\documentstyle[times,graphics,astrobib,amssymb,psfig,epsfig]{mn2e}

\newcommand{\de}{{\rm d}}
\newcommand{\bea}{\begin{eqnarray}}
\newcommand{\eea}{\end{eqnarray}}
\newcommand{\f}{\frac}

\title[Star formation, luminosity function and reionization]
{Probing the star formation history using the redshift evolution of luminosity functions}
\author[Samui, Srianand \& Subramanian] 
{Saumyadip Samui\thanks{E-mail: samui@iucaa.ernet.in},
Raghunathan Srianand\thanks{E-mail: anand@iucaa.ernet.in},
Kandaswamy Subramanian\thanks{E-mail: kandu@iucaa.ernet.in} \\
IUCAA, Post Bag 4, Ganeshkhind, Pune 411 007, India.}

\begin{document}

\maketitle
\begin{abstract}
We present a self-consistent, semi-analytical $\Lambda$CDM model of star
formation and reionization. For the cosmological parameters favored by
the WMAP data, our models consistently reproduce the electron scattering
optical depth to reionization, redshift of reionization and the observed
luminosity functions (LF) and hence the star formation rate (SFR) density 
at $3\le z \le6$ for a reasonable range of model parameters. While simple
photoionization feedback produces the correct shape of LF at $z = 6$, for
$z = 3$ we need additional feedback that suppresses star formation activities
in halos with $10^{10}\lesssim  (M/M_\odot)\lesssim 10^{11}$. Models with
prolonged continuous star formation activities are preferred over those with
short bursts as they are consistent with the existence of a Balmer break in 
considerable fraction  of observed galaxies even at $z\sim 6$. The halo number
density evolution from the standard $\Lambda$CDM structure formation model
that fits LF up to $z=6$ is consistent with the upper limits on $z\simeq 7$ LF
and source counts at $8\le z\le12$ obtained from the Hubble Ultra
Deep Field (HUDF) observations without requiring any dramatic change in the 
nature of star formation. However, to reproduce the observed LF at $6\le z \le 10$,
obtained from the near-IR observations around strong lensing clusters, we need
a strong evolution in the initial mass function, reddening correction and 
the mode of star formation at $z\gtrsim 8$. We show that low mass molecular
cooled halos, which may be important for reionizing the universe, are not
detectable in the present deep field observations even if a considerable
fraction of its baryonic mass goes through a star burst phase. However, their
presence and contribution to reionization can be inferred indirectly from the
redshift evolution of the luminosity function in the redshift range
$6\le z\le 12$. In our model calculations, the contribution of low mass halos
to global SFR density prior to reionization reveals itself in the form of
second peak at $z\ge 6$.  However this peak will not be visible in the
observed SFR density as a function of $z$ as most of these galaxies have
luminosity below the detection threshold of various ongoing deep field surveys.
Accurately measuring the LF at high redshifts can be used to understand the
nature of star formation in the dark ages and probe the history of reionization.

\end{abstract}
 
\section{Introduction}

Understanding how and when the dark ages ended and led to the reionization of
the intergalactic medium (IGM) is one of the holy grails of modern cosmology.
The collapse of the first non-linear structures and the associated star
formation possibly provide the first sources of UV photons which ionized
the IGM. Direct observations of these earliest star forming `galaxies' is
rapidly on the rise, with data constraining the luminosity functions of
galaxies and hence the redshift evolution of star formation rate (SFR) density
till $z\simeq10$. At the same time tight constraints are being set on the
epoch of reionization by studying spectra of the highest redshift QSOs and the
ongoing WMAP satellite observations of the Cosmic Microwave Background
(CMB) polarization.  It is imperative to construct models of structure
formation that explain the wealth of available data, and so probe the nature
of star formation in the dark ages as well as in the post reionization era.
This forms the basic motivation of the present work.

The absence of the Gunn-Peterson absorption (Gunn \& Peterson, 1965)
blue-ward of the Lyman-$\alpha$ emission from background QSOs
had indicated that the IGM is highly ionized at least up to
redshifts of about $5$ or so. However, recent detections of a strong
Gunn-Peterson trough in the spectra of QSOs with $z\simeq6$ and limits
obtained on the sizes of ionized regions around the highest $z$ QSOs,
indicate a significantly neutral IGM above $z \sim 6$ \cite{loeb,fan}.
The CMB observations are still intriguing, with a recent downward
revision of the the optical depth to electron scattering, from
a $\tau_e =0.17^{+0.08}_{-0.07}$  based on the first year WMAP
data \cite{spergel03}, to a value $\tau_e = 0.09^{+0.03}_{-0.03}$ from the
three year data \cite{spergel06}. The latter measurement would naively
suggest a reionization redshift $7 \le z_{re} \le 12$. This is consistent
with that indicated by the quasar observations.
However, WMAP 3rd year data also indicate a lower power in
density fluctuations  (with $\sigma_8 = 0.74^{+0.05}_{-0.06}$) and also a
redder power spectrum (with a scalar spectral index $n_s =0.952^
{+0.015}_{-0.019}$). Both these effects decrease the predicted
number density of collapsed objects, and hence can make it potentially
difficult to explain the observed $\tau_e \sim 0.09$ (Alvarez et al. 2006).

Direct measurements of luminosity function and hence the SFR density
up to the redshift that is consistent with the $z_{re}$ suggested by the
WMAP data is now possible, thanks to the photometric dropout techniques
(e.g. Steidel et al. 2003). For $z\lesssim 6$ we have several sets of
observations from different groups which  appear to be in reasonably good
agreement with each other \cite{iwata03,sawicki,bouwens06}. These observations
put tight constraints on luminosity functions at these epochs and hence on SFR
density.  However observational constraints on the luminosity function for
redshifts $z>6$ are scanty and have more uncertainty than the $z\lesssim6$
data. Bouwens et al. (2005), based on the NICMOS-UDF data reported a rapid
decline in the SFR density at $z\gtrsim 5$. In striking contrast, the
SFR density estimated from Ly$\alpha$ emitters detected at $z  = 5.7$
and $6.5$ implies no substantial decrease with $z$ \cite{hu06}.
The decline in the SFR density is also not supported by near-IR observations
of Richard et al. (2006) around lensing clusters, obtained with VLT/ISAAC.
Selection biases and cosmic variance could provide possible reasons for the
difference \cite{hu06}. As we will show here, clarifying this issue
observationally is of paramount importance to probe the nature of the first
star formation.

Reionization also feeds back on star formation by suppressing the collapse of
the gas into low mass halos \cite{tw96}. It is then important to model the
redshift evolution of luminosity function, SFR density and reionization
simultaneously, in a self-consistent manner, taking account of such radiative
feedback. We do this here adopting a semi-analytical approach. This is also
motivated by the need to explore the sensitivity to model parameters in an
extensive fashion.

In section 2 we outline our semi-analytic models for star formation and
reionization and discuss how to compute the redshift evolution of the
luminosity function and integrated source counts for high-$z$ galaxies.
The resulting reionization history is described in section 3, while sections 4 
and 5 focus on the results for the UV luminosity function of high redshift
galaxies. We elaborate further in section 6, the utility of the $z\ge 6$
luminosity functions in probing reionization history. The redshift evolution of
the SFR density inferred from our models is presented in section 7, and a
discussions of results and our conclusions are presented in the last section.
In most of this work we use the cosmological parameters consistent with the
recent WMAP data ($\Omega=1$, $\Omega_m = 0.26$, $\Omega_\Lambda = 0.74$,
$\Omega_b=0.044$, $h = 0.71$, $\sigma_8=0.75$ and $n_s =0.95$).

\section{Semi-analytic models}
\label{semianalytics}

\subsection{Redshift evolution of star formation}

The formation and evolution of galaxies and the associated star formation
histories have been studied extensively using both numerical simulations and
semi-analytic models \cite{chiu,tirth,springel03,nagamie06}. Here, we use the
modified Press-Schechter (PS) formalism of Sasaki (1994) to study the redshift
evolution of the SFR density (see also Chiu \& Ostriker (2000); Choudhury \&
Srianand (2002)). In this formalism the number density of collapsed objects
having mass in the range $(M, M + \de M)$, which are formed at the
redshift interval $(z_c, z_c + \de z_c)$ and survive till redshift
$z$ is, \cite{sasaki94,chiu}
\bea
N(M,z,z_c)~ \de M~ \de z_c &=& N_M(z_c) \left(\f{\delta_c}{D(z_c) 
\sigma(M)}\right)^2 \f{\dot{D}(z_c)}{D(z_c)}\;\nonumber \\
& \times & \f{D(z_c)}{D(z)} \f{\de z_c}{H(z_c) (1 + z_c)}~ \de M.
\label{eqnmPS}
\eea
Here, the overdot represents time derivative, $N_M(z_c)~ \de M$ is the number
of collapsed objects per unit comoving volume within a mass range
$(M, M+\de M)$ at redshift $z_c$, known as the PS mass function \cite{ps74},
and $\delta_c$ is the critical over density for collapse, usually taken to be
equal to $1.686$ for a matter dominated flat universe $(\Omega_m=1)$.
This parameter is quite insensitive to cosmology and hence the same value can
be used for all cosmological models \cite{ecf96}. Further, $H(z)$ is the
Hubble parameter, $D(z)$ the growth factor for linear perturbations and
$\sigma(M)$ the rms mass fluctuation at a mass scale~$M$. The ratio
$D(z_c)/D(z)$ gives the probability that a halo which collapsed at $z_c$
survives till $z$.
Note that $N(M,z,z_c)$ is the formation rate of halos weighted by their
survival probability, and integrating it over $z_c$ from $z$ to $\infty$ 
gives the PS mass function $N_M(z)$ at any redshift $z$:
that is $N_M(z) = \int_z^\infty N(M,z,z_c)~ \de z_c$.

Next, we assume that the SFR at $z$ of a halo of mass $M$ that has collapsed
at an earlier redshift $z_c$, is given by \cite{chiu,tirth}
\bea
\dot{M}_{\rm SF}(M,z,z_c) &=& f_{*} \left(\f{\Omega_b}{\Omega_m} M \right) 
\f{t(z)-t(z_c)}{\kappa ^2~ t_{\rm dyn}^2(z_c)} \nonumber \\
& &  \times \exp\left[-\f{t(z)-t(z_c)}{ \kappa ~t_{\rm dyn}(z_c)}\right].
\label{eqnsf}
\eea
Here, $f_{*}$ is the fraction  of total baryonic mass in a halo that will be
converted to stars. The function $t(z)$ gives the age of the universe at
redshift $z$; thus, $t(z)-t(z_c)$ is the age of the collapsed halo at $z$.
$t_{\rm dyn}$ is the dynamical time-scale and given by \cite{chiu,bl01} 
\bea
t_{\rm dyn}(z) &=& \sqrt{\f{3\pi}{32 G \rho _{\rm vir}(z)}}.
\eea
Here,
\bea
\rho_{\rm vir}(z) &=& \Delta_c(z) \rho_c(z) \nonumber \\
\Delta_c (z) &=& 18 \pi ^2 + 82 d(z) - 39 d^2(z) \nonumber \\
d(z) &=& \f{\Omega_m ( 1 + z )^3}{\Omega_m ( 1 + z )^3 + \Omega_{\Lambda}} - 1 \nonumber \\
\rho_c(z) &=& \f{3 H^2(z)}{8 \pi G}. \nonumber 
\eea
The duration of star formation activity in a halo depends on the value of
$\kappa $. For a given value of $\kappa$ star formation occurs in a continuous
mode with a peak at $\kappa ~t_{\rm dyn}$ and decaying  exponentially
afterward. Note that $\kappa \to 0$ corresponds to the star formation
occurring in a single burst.

We can then calculate the cosmic SFR per unit comoving volume at a redshift
$z$ using,
\begin{equation}
\dot{\rho}_{\rm SF}(z) = \int\limits_z^{\infty} \de z_c 
\int\limits_{M_{\rm low}}^{\infty} \de M' \dot{M}_{\rm SF}(M',z,z_c) 
\times N(M',z,z_c).
\label{eqnsfr}
\end{equation}
The lower mass cutoff ($M_{\rm low}$) at a given epoch is decided by the
cooling efficiency of the gas and different feedback processes. In the
absence of UV background radiation $M_{\rm low}$ is decided
only by the cooling efficiency of the gas.
In the early universe, recombination line cooling from hydrogen and helium
lines are favored as the heavier elements are absent. However, such cooling
is effective only above temperatures of about $10^4$ K. Thus gas in halos
with virial  temperatures in excess off $10^4$ K can cool and collapse to
form stars. However, if one can increase the H$_2$ content of the gas then
molecular line cooling can lead to the formation of cold gas condensations
within the low mass halos \cite{tsrbap97}.
Haiman, Abel \& Rees (2000) have shown that such cooling can be efficient for
$T_{\rm vir}\ge300$ K. The presence of Lyman and Werner band photons that are
produced by luminous objects can easily destroy these H$_2$ molecules. So
survival of star formation activities in low mass halos is always very
uncertain. In what follows we consider models with  $M_{\rm low}$
corresponding to a virial temperature, $T_{\rm vir} = 10^4$~K (hereafter
``atomic cooling model'') and $300$~K (``molecular cooling model'').

Ionization of the IGM by UV photons enhances the temperature of the gas
thereby increasing the Jean's mass for the collapse. Thus, $M_{\rm low}$ is
increased in the ionized regions due to photoionization heating. It is known
from simulations that the photoionizing background suppresses galaxy formation
within halos with circular velocities ($v_c$) less than about $35$~km~s$^{-1}$,
while the mass of cooled baryons is reduced by 50\% for  halos with circular
velocities $\sim 50$~km~s$^{-1}$ \cite{tw96}. However, the exact value will
depend on the intensity and spectral shape of the ionizing background
radiation. Therefore, the cutoff in $v_c$ can in principle be redshift
dependent \cite{benson,dijkstra}. 
In the ionized fraction of the universe, our models assume complete
suppression of star formation in halos below circular velocity
$v_c=35$~km~s$^{-1}$ and no suppression above circular velocity of
$95~$km s$^{-1}$. For intermediate masses, we adopt a linear fit  from $1$
to $0$ for the suppression factor (as in Bromm \& Loeb, 2002). The ionized
fraction of the universe itself, at a given epoch, is computed using the
simple model of reionization described below.

Semi-analytical $\Lambda$CDM models of galaxy formation without feedback
usually also overproduce the number of high luminosity galaxies compared
to the observations (see Somerville \& Primack, 1999). Further, recent
observations of high-$z$ galaxies suggest that the SFR in massive galaxies was
higher at high-$z$ (i.e. $z\simeq2$) compared to that in the local universe,
contrary to the naive predictions of hierarchical structure formation
(Cowie et al. 1996; Juneau et al.  2005; Croton et al 2006). Thus, we need to
incorporate the declining star formation activities in the massive halos.
This can arise, physically, due to the effect of AGN feedback (Bower et al.
2005; Best et al. 2006).
For simplicity we model this by multiplying the
integrand in Eq.~(\ref{eqnsfr}) with a suppression factor
\begin{equation}
f_{\rm sup} = \f {1}{ 1 + \left(\f {M}{10^{12} M_\odot }\right)^3}
\label{eqnsup}
\end{equation}
which sharply decreases star formation in high mass halos above
$10^{12} M_\odot$. The main conclusions of our work are insensitive to the
exact nature of this cutoff as long as the typical mass above which star
formation is suppressed is not much smaller than $10^{12} M_\odot$.

\subsection{ Redshift evolution of the luminosity function}

Although many works emphasize the measurements of SFR density at different
redshifts, the more directly observed quantity is rather the luminosity
function. The SFR density is merely a conversion from the measured luminosity
function to the star formation rate assuming a continuous star formation
with some initial mass function (IMF) for the formed stars. So a more accurate
comparison of theory with observations will be possible if we compare the
luminosity function at different epochs computed from our semi-analytical
models with the observed luminosity functions. Furthermore, very low mass
halos could contribute significantly to the theoretically computed SFR density,
but not at all to the detected individual galaxies (and hence the observed
luminosity function or the inferred SFR density). Therefore it is important to
calculate the luminosity functions predicted by our semi-analytic models to
constrain parameters of our models more accurately.

\begin{figure}
\centerline{
\psfig{figure=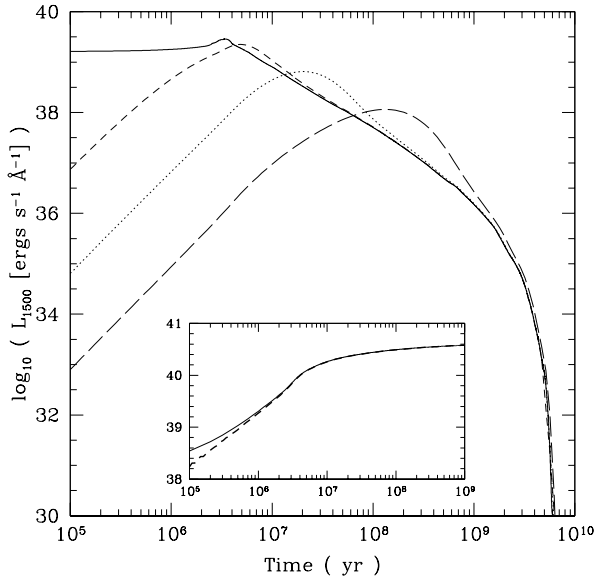,width=8.0cm,angle=0.}}
\caption[]{Luminosity at $1500~$\AA  ~of a halo with stellar mass $10^6 ~M_\odot
$ collapsed at $z_c = 10$. The IMF is Salpeter from $1 - 100~ M_\odot$ with 
metallicity $Z=0.0004$. The solid curve is for burst model (obtain from
``Starburst99''). Other three curves are obtained by using Eq.~(\ref{eqnlf}) and 
the SFR given by Eq.~(\ref{eqnsf}) for $\kappa = 1$ (long-dashed),
$1/10$ (dotted) and $1/100$ (short-dashed). In insert, we show the luminosity
of a galaxy undergoing a continuous star formation at a rate of $1~M_\odot$
yr$^{-1}$ for comparison. The solid curve is the result from ``Starburst99''
code (Fig. 54 of original 1999 dataset) and dashed curve is our model
prediction. The IMF is Salpeter in the mass range $1-100~M_\odot$ and the
metallicity is $Z=0.001$. 
}
\label{figlumcomp}
\end{figure}

The luminosity function is computed as follows. From ``Starburst99" code
\footnote{http://www.stsci.edu/science/starburst99} \cite{leith99} we obtain 
$l_{1500}(t)$, the luminosity at $1500~$\AA~
as a function of time, produced for every solar mass being converted to stars
in a single (instantaneous) starburst. This quantity depends mainly on the IMF
and metallicity of the gas. In our model, star formation is a continuous
process lasting for a few dynamical time. We  therefore compute the luminosity
of a galaxy of age $T$ using
\begin{equation}
L_{1500} (M,T) = \int\limits_{T}^0 \dot{M}_{\rm SF}(M,T - \tau) \,
		l_{1500}(\tau)~ \de\tau.
\label{eqnlf}
\end{equation}
Here, the age of the galaxy formed at $z_c$ and observed at $z$ is 
$T(z,z_c) = t(z) - t(z_c)$. As a check on our prescription for calculating the
luminosity as function of time, we reconstruct the luminosity evolution of a
galaxy undergoing a constant star formation using convolution integral in
Eq.~(\ref{eqnlf}) and compare it with the luminosity evolution directly
obtained from ``Starburst99'' code. This comparison, shown in
Fig.~\ref{figlumcomp} as an insert, demonstrates that
our prescription could provide a fairly accurate representation
of the luminosity expected from models with continuously varying SFR.
In Fig~\ref{figlumcomp} we show the luminosity at $1500~$\AA~  calculated
with our prescription, from a galaxy with a variable SFR.
The galaxy is assumed to collapse at $z_c=10$ and form 
$10^6~M_\odot$ of stars with a Salpeter IMF (and masses between
$1 - 100~ M_\odot$), at a rate given by Eq.~(\ref{eqnsf}), and with 
$\kappa =1,~1/10$ and $1/100$. As per our expectation, when 
$\kappa$ becomes small (say $\kappa =1/100$)  the luminosity at 
$1500~$\AA~ as a function of time is close to that expected 
from a single star burst. It is also evident from Fig.~\ref{figlumcomp}
that the peak luminosity is higher and occurs earlier
for lower values of $\kappa$. 
%

Hence, for any given halo of mass $M$ which collapses at $z_c$, and undergoes
star formation as given by Eq.~(\ref{eqnsf}), one can compute its luminosity
evolution. The luminosity can be converted to a standard absolute magnitude,
say the AB magnitude using
\begin{equation}
M_{AB} = -2.5\log _{10}(L_{\nu 0}) + 51.60
\label{eqnmagtolf}
\end{equation}
where the luminosity is in units of erg s$^{-1}$ Hz$^{-1}$ (Oke \& Gunn 1983). 
The luminosity function $\Phi(M_{AB}, z )$ at any redshift $z$ is then given by
\begin{eqnarray}
& &\Phi(M_{AB}, z ) ~\de M_{AB}  \nonumber \\
& &\quad\quad = \int\limits_z^\infty \de z_c~ N(M,z,z_c) 
\:\:\:\:\:\f{\de M}{\de L_{1500}}~\f{\de L_{1500}}{\de M_{AB}} ~ \de M_{AB}
\label{eqnlfmag}
\end{eqnarray}
where $N(M,z,z_c)$ is given by Eq.~(\ref{eqnmPS}). In what follows, we will be
comparing the high redshift luminosity functions computed using
Eq.~(\ref{eqnlfmag}) with observations.
Further, one can directly integrate the theoretically computed luminosity
function at a given $z$ to a given luminosity limit, fold in a conversion
factor between the luminosity and the SFR to obtain the SFR density at that
redshift. Note even though Eq.~(\ref{eqnsf}) directly gives the SFR density,
it is better to obtain it by integrating the luminosity function.
This is because the SFR density is usually measured by integrating
the observed luminosity function above some luminosity threshold (for example
$0.3L^*_{z=3}$). This issue is dealt with in greater detail in section 7.

\subsection{ Integrated source counts}
For redshifts $z\gtrsim 8$, Bouwens et al. (2005) give upper limits on the
number of sources detected up to the limiting apparent magnitude that their
observations reach for $8<z<12$. They have then used these upper limits to set
limits on the the SFR density at $z\gtrsim 8$. In order to compare with such
observation at high redshifts we also compute the integrated source counts as
a function of the limiting apparent magnitude. The number of galaxies per unit
solid angle in the sky per unit redshift interval with apparent magnitude less
than a limiting magnitude $m_0$ is given by \cite{peebles,paddy}
\begin{eqnarray}
\f{\de n(z,m<m_0)}{\de \Omega \de z} = {\cal N}(z,m<m_0) r^2_{\rm em}(z) d_H(z)
\label{eqnsc}
\end{eqnarray}
where
$$
d_H(z) = \f{c}{H_0\left[ \Omega_{\lambda} + \Omega_m (1 +z)^3 \right]^{1/2}}
$$
is the `Hubble Radius' and
$$
r_{\rm em}(z) = c\int\limits_0^z \f{\de z}{H_0\left[ \Omega_{\lambda} + \Omega_m (1 +z)^3 \right]^{1/2}}.
$$
Further, ${\cal N}(z,m<m_0)$ is the comoving number density of object at
redshift $z$ having apparent magnitude less than $m_0$, i.e.
\begin{equation}
{\cal N}(z,m<m_0) = \int \limits_{-\infty}^{M_0(z,m_0)} \Phi _{M_{AB}}(M_{AB}, z )
~\de M_{AB}
\end{equation}
The relation between apparent magnitude and absolute magnitude is given by \cite{peebles,paddy}
\begin{equation}
m-M=25 + 5\log_{10}[3000 (1+z) r_{\rm em}(z) H_0/ c] - 5\log_{10}~h.
\end{equation}

To give a rough idea of the numbers involved, suppose a $10^{10} M_\odot$ dark
matter halo collapses at $z_c = 10$ and undergoes a burst of star formation,
with a fraction $f_*=0.5$ going into stars having a Salpeter IMF from
$1 - 100 ~M_\odot$. Let us assume that the metallicity of the gas is
$Z=0.0004$. Then the luminosity at early epochs (up to a few Myrs) is
$\sim 1.4 \times 10^{42}$ ergs s$^{-1}$ \AA$^{-1}$, corresponding to an 
absolute AB magnitude of $-23.46$ (with no extinction correction) or $-21.82$
(with an extinction correction factor of $4.5$). The corresponding apparent
magnitudes are $26.7$ and $28.34$ respectively. In the survey by Bouwens et al.
the limiting magnitude was $\sim 28.5$. So the detection of these halos with
present technology depends on the amount of dust reddening in these galaxies.
If the galaxy under goes a continuous star formation then its luminosity
can be much smaller, making it more difficult to detect.

\subsection{Cosmological reionization}

\begin{table}
\begin{center}
\caption{Values of $n_\gamma$ for different model parameters$^1$.}
\begin{tabular}{c c c c }
\hline
$m_{\rm low} (M_\odot)$ & $m_{\rm up} (M_\odot)$ & Metallicity($Z$) & $n_\gamma $ \\ \hline
1 & 100 & 0.040 & 4800 \\
1 & 100 & 0.020$^\dag$ & 5675  \\
1 & 100 & 0.008 & 6530 \\
1 & 100 & 0.004 & 7245 \\
1 & 100 & 0.001 & 8710 \\
1 & 100 & 0.0004 & 10450 \\
0.5 & 100 & 0.0004 & 7780 \\
0.1 & 100 & 0.0004 & 4100 \\
10 & 100 & 0.0004 & 33810 \\
50 & 500 & $10^{-7}$ & 83000$^\ddag$ \\ 
\hline
\multicolumn{4}{l}{$^1$All the models use Salpeter IMF with $\alpha = 2.35$.}\\ 
\multicolumn{4}{l}{$^\dag$ Solar metallicity ($Z_\odot$). }\\
\multicolumn{4}{l}{$^\ddag$ Taken from Schaerer (2003).}
\end{tabular}
\label {ngammatable}
\end{center}
\end{table}

In order to calculate the radiative feedback at different redshifts
we need to know the ionization history of the universe. For this 
purpose, we consider the following simple model of reionization 
that assumes
(i) all the baryons in the IGM are in the form of  hydrogen and (ii) all
the Lyman continuum photons that escape a star forming galaxy are used for
reionizing the IGM. The fraction of ionized hydrogen ($f_{HII}$) evolves
as \cite{bl01},
\bea
\f {\de f_{HII}}{\de z } & =& \f {\dot{N}_{\gamma}}{n_H(z)}~\f{\de t}{\de z}
 ~-~\alpha_B n_H(z)f_{HII} C~\f{\de t }{\de z}. 
\eea
Here, $\dot{N}_{\gamma}$ is the rate of UV photons escaping into the IGM and
$n_H(z)$ is the proper number density of the hydrogen atoms. The clumping factor
of the IGM, $C$, is defined as $C\equiv \langle n^2_H \rangle / {\bar n}_H^2$
and $\alpha_B$ is the case B recombination coefficient, at
$T = 3 \times 10^4$ K. The first term on the right is the rate of ionization
and second term is rate of recombination, weighted by the $f_{HII}$, 
as recombinations take place only in the ionized region.
${\dot N}_{\gamma}$ is obtained from the SFR density using,
\bea
\dot{N}_{\gamma} = \f{\dot{\rho}_{\rm SF}(z)(1+z)^3}{m_p} n_\gamma f_{esc} .
\eea
Here $n_\gamma$ is the number of ionizing photons released per baryon  
of stars formed and $f_{esc}$ is the fraction of these photons which 
escape from the star forming halo. The value of $n_\gamma$ depends on
the IMF of the forming stars. For a Salpeter IMF (with 
$m_{\rm low}= 0.1~M_{\odot}$, $ m_{\rm up}=100~M_{\odot}$) $n_\gamma
\sim 4000$. However,  for first generation of metal free stars, the IMF could
be biased towards very massive stars. This can give much larger values 
for $n_\gamma \sim 80000$  \cite{schaerer,haiman06}. In Table~\ref{ngammatable}
we summarize the values of  $n_\gamma$ for different IMFs and metallicities
obtained from `Starburst99' that are used in our subsequent calculations.

Clearly $f_{HII}$ as a function of $z$ depends on our choice of 
$n_\gamma$, $f_{esc}$ and $C$.  In what follows, we use $f_{esc}=0.1$ and
$n_\gamma$ corresponding to the assumed IMF. For the clumping factor $C$,
we have assumed the following simple form given by \cite{haiman06}
\begin{equation}
C(z) = 1 + 9\left(\f{7}{1+z}\right)^2
\end{equation}
for $z\ge6$ and $C= 10$ for $z<6$. 
We also compute the electron scattering optical depth ($\tau_e$) in order
to compare it with the recent WMAP observation. 

Having established the basic framework of semi-analytic models, in the
following sections we present our self-consistent results of reionization,
luminosity function and SFR density.

\section {Reionization history in different models}

\begin{table}
\caption[]{ Results of reionization for atomic cooling models:$^\dag$
}
\begin{center}
\begin{tabular}{c c c c c c c}
\hline
$m_{\rm low}$ ($M_\odot$) &$Z$  &
$\sigma_8 $ & $n_s$ & $\kappa$ & $z_{re}$ & $\tau _e$ \\ \hline
0.1 &  0.0004 & 0.75 & 0.95 & 1.00 & 5.9 & 0.066 \\
0.5 &  0.0004 & 0.75 & 0.95 & 1.00 & 6.6 & 0.076 \\
  1 &  0.0004 & 0.75 & 0.95 & 1.00 & 7.0 & 0.080$^a$ \\
10 &  0.0004 & 0.75 & 0.95 & 1.00 & 8.4 & 0.097$^f$ \\ \hline
 1 & 0.001 & 0.75 & 0.95 & 1.00 & 6.8 & 0.077 \\
 1 & 0.004 & 0.75 & 0.95 & 1.00 & 6.7 & 0.075 \\
 1 & 0.008 & 0.75 & 0.95 & 1.00 & 6.4 & 0.073 \\
 1 & 0.020 & 0.75 & 0.95 & 1.00 & 6.3 & 0.071 \\
 1 & 0.040 & 0.75 & 0.95 & 1.00 & 6.0 & 0.069$^b$ \\ \hline
 1 & 0.0004 & 0.85 & 0.95 & 1.00 & 8.5 & 0.100 \\
 1 & 0.0004 & 0.95 & 0.95 & 1.00 & 9.7 & 0.122 \\
 1 & 0.0004 & 0.75 & 1.00 & 1.00 & 7.8 & 0.093 \\
 1 & 0.0004 & 0.85 & 1.00 & 1.00 & 9.3 & 0.117 \\
 1 & 0.0004 & 0.95 & 1.00 & 1.00 & 10.9 & 0.142$^c$ \\ \hline
 1 & 0.0004 & 0.75 & 0.95 & 0.50 & 7.2 & 0.088 \\
 1 & 0.0004 & 0.75 & 0.95 & 0.33 & 7.0 & 0.091 \\
 1 & 0.0004 & 0.75 & 0.95 & 0.25 & 7.1 & 0.092 \\
 1 & 0.0004 & 0.75 & 0.95 & 0.20 & 7.0 & 0.093 \\
 1 & 0.0004 & 0.75 & 0.95 & 0.11 & 7.0 & 0.095$^d$ \\ \hline
 10 & 0.0004 & 0.75 & 0.95 & 0.50 & 8.8 & 0.105 \\
 10 & 0.0004 & 0.75 & 0.95 & 0.25 & 8.3 & 0.111 \\
 10 & 0.0004 & 0.75 & 0.95 & 0.11 & 8.4 & 0.114$^e$ \\
\hline
\multicolumn{7}{l}{$^\dag$All the models assume $m_{\rm up} = 100~M_\odot$, $f_* = 0.5$ and $f_{esc}=0.1$.}\\
\multicolumn {7}{l}{$^a$model A; $^b$model B;  $^c$model C; $^d$model D; $^e$model E; $^f$model F.}
\end{tabular}
\end{center}
\label{tab_reion}
\end{table}

The epoch of reionization and hence the electron scattering optical depth,
$\tau_e$, are sensitive to cosmological parameters, mode of star formation and
escape fraction of UV photons, $f_{esc}$. We define $z_{re}$ as the redshift
when the ionized fraction $f_{HII}$ becomes unity.
In Table~\ref{tab_reion} we show $z_{re}$ and $\tau_e$ for a range of
parameters considered in our study keeping $f_{esc}=0.1$, $f_*=0.5$ and
taking account of only the atomic cooled halos. Note $f_*=0.5$ used here is
constrained by the observed luminosity functions discussed in the following
section. The recent WMAP data gives $\tau_e = 0.09\pm0.03$. Models which
assume Salpeter IMF with $m_{\rm low}\le 1~M_\odot$, $\kappa = 1$ and a range
of metallicities produce $\tau_e$ in the lower end of the allowed range from
the WMAP 3rd year data. However, models considering star formation with a
top-heavy IMF and adopting a lower value of $\kappa$ produce slightly higher
values of $\tau_e$.

\begin{figure}
\centerline{
\psfig{figure=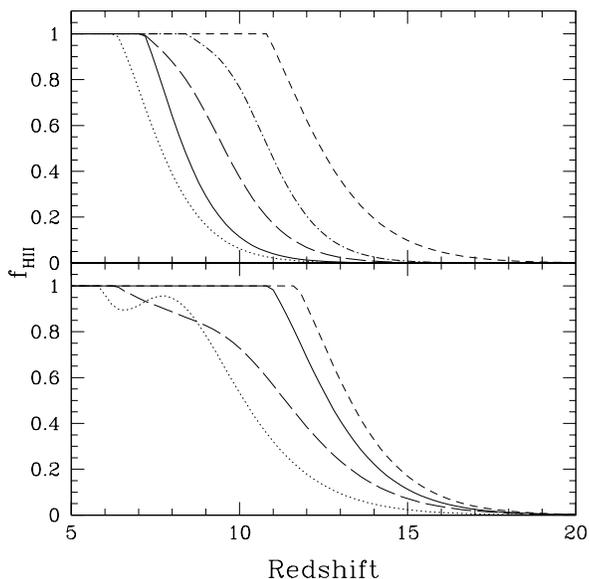,height=8cm,width=8cm,angle=0.0}}
\caption[]{The reionization history for some atomic cooling 
(top panel) and molecular cooling (bottom panel) models.
{\it Top panel:} The solid, dotted, short-dashed, long-dashed and dot-dashed 
curves are for models A, B, C, D and E respectively (see Table~\ref{tab_reion}).
{\it Bottom panel:}  The solid, dotted, short-dashed and long-dashed 
curves are for models M$_1$, M$_2$,  M$_3$ and M$_4$ respectively (see Table~\ref{tab_reion_m}).
}
\label{fig_reion}
\end{figure}

Therefore, atomic cooling models with cosmological parameters constrained by 
the 3rd year WMAP data ($\sigma_8 =0.75$ and $n_s=0.95$) produce
consistent values of $\tau_e$  for a range of star formation scenarios. 
Also the inferred reionization redshifts are consistent with observations of 
the highest redshift QSOs (Fan et al. 2006) and Lyman-$\alpha$ emitters
(Iye et al. 2006). Reionization histories for some of these models are shown
in top panel in Fig.~\ref{fig_reion}.
From Table~\ref{tab_reion} it is also clear that the models with higher values
of $m_{\rm low}$, $\sigma_8$ and $n_s$ produce reionization at slightly higher
redshifts with higher $\tau_e$. Further, for a given value of $f_*f_{esc}$ the
redshift of reionization only depends weakly on $\kappa$ as integrated
star formation and hence the total number of UV photons escaping a
halo remains the same. However, $\tau_e$ is larger for smaller
$\kappa$ as most of the star formation in halos occur over a shorter time-scale
(see Fig.~\ref{figlumcomp}) thereby establishing H~{\sc ii} regions very quickly.
\begin{table}
\caption{Results of reionization for molecular cooling models.$^{1,2,3}$}
\begin{center}
\begin{tabular}{c c c c c c c}
\hline
m$_{\rm low} (M_\odot)$ &m$_{\rm up} (M_\odot)$ & $n_\gamma$ & $f_*$ & $\kappa$ & $z_{re}$ & $\tau_e$ \\ \hline
 1 & 100 & 10450 & 0.50 & 1 & 10.8 & 0.145$^a$ \\ 
 1 & 100 & 10450 & 0.10 & 1 &  5.9 & 0.105$^b$ \\   
50 & 500 & 83000 & 0.50 & 1 & 14.2 & 0.194 \\ 
50 & 500 & 83000 & 0.10 & 1 & 11.6 & 0.155$^c$ \\ 
10 & 100 & 33800 & 0.50 & 1 & 12.6 & 0.171 \\ 
10 & 100 & 33800 & 0.10 & 1 & 10.0 & 0.134 \\ \hline
 1 & 100 & 10450 & 0.50 &1/2 & 6.3 & 0.154 \\ 
 1 & 100 & 10450 & 0.50 &1/4 & 6.4 & 0.162 \\ 
 1 & 100 & 10450 & 0.10 &1/2 & 6.2 & 0.114 \\ 
 1 & 100 & 10450 & 0.10 &1/4 & 6.2 & 0.121$^d$ \\ 
10 & 100 & 33800 & 0.10 &1/2 & 6.4 & 0.143 \\ 
10 & 100 & 33800 & 0.10 &1/4 & 6.5 & 0.150 \\
50 & 500 & 83000 & 0.10 &1/2 & 7.8 & 0.166 \\
50 & 500 & 83000 & 0.10 &1/4 & 6.9 & 0.174 \\ 
\hline
\multicolumn{7}{l}{$^1$All models assume $f_{esc} = 0.1$, $\sigma_8 = 0.75$ and $n_s = 0.95$.}\\
\multicolumn{7}{l}{$^2$Atomic cooled halos have parameters similar to model A in Table~\ref{tab_reion}.}\\
\multicolumn{7}{l}{$^3$$\kappa$ as given in the table is used for all the halos.}\\
\multicolumn {7}{l}{$^a$model M$_1$; $^b$ model M$_2$;  $^c$ model M$_3$; $^d$ model M$_4$.}
\end{tabular}
\end{center}
\label{tab_reion_m}
\end{table}

%
Now consider the effect of star formation in molecular cooled halos. Molecular
cooled halos were proposed as a main source for early reionization in order to
reproduce the high optical depth reported from the 1st year WMAP data.
In Table~\ref{tab_reion_m} we have shown the results when star formation is
also allowed in such halos. Reionization histories for some of these models
are shown in bottom panel in Fig.~\ref{fig_reion}. It can be noted that if we
use $f_*=0.5$ and $f_{esc} =0.1$ also in the case of molecular cooled halos
(model M$_1$) the resulting optical depth is higher than the value obtained
 from the WMAP 3rd year data.
It is obvious from the table that inclusion of star formation in molecular cooled 
halos increases the value of $\tau_e$. However, this need  not always leads to
a higher value of $z_{re}$. This happens because we self-consistently calculate
the reionization history where the radiative feedback suppresses the star formation
in smaller mass halos. Such an effect is very clear for model M$_2$. Without star
formation in molecular cooling halos we had $z_{re} = 7.0$ and $\tau_e =0.080$ 
(see model A in Table~\ref{tab_reion}). Inclusion of star formation in molecular
cooled halos has increased the value of $\tau_e$ to $0.105$ but decreased
$z_{re}$ to $5.9$.

From Table~\ref{tab_reion_m} one can conclude that to obtain $\tau_e$ within 
the $1\sigma$ value predicted by WMAP 3yr data star formation
in molecular cooled halos should be in a continuous mode with normal Salpeter
IMF if we consider the same efficiency factors as atomic cooled halos (i.e. $f_*=0.5$ and $f_{esc} =0.1$). 
The models with top-heavy IMF will produce consistent reionization only
when we reduce either of these two efficiencies drastically. 
For example, a Pop III mode of star formation in molecular cooled halos,
with $n_\gamma \sim 83,000$ leads to a $\tau_e \sim 0.17$ for
$f_*f_{esc} = 0.01$, and so exceeds the $\tau_e$ inferred from the
WMAP 3rd year data at a $2\sigma$ level. Clearly based on the constraints on 
reionization from WMAP data alone it is not possible to independently
constrain both $f_*$ as well as $f_{esc}$. We can only constrain the product
$f_* f_{esc}$, and from Table~\ref{tab_reion_m}, it appears that
one needs this product to be at least smaller than $\sim 0.01$ for a top-heavy IMF.
Thus the recent WMAP observations are better 
consistent with a low efficiency of the molecular cooled
halos in reionizing the universe. This is in consonance with the results of
Choudhury \& Ferrara (2006), Haiman \& Bryan (2006) and Greif \& Bromm (2006).

We now have a set of models that can produce a $\tau_e$ consistent with the
3rd year WMAP data. However, as we will show in the next section, the
luminosity functions and hence the global star formation rate density in
these models can be very different. Therefore, one can use the observed
luminosity functions at different epochs to get better 
constraints on our model parameters. We do this in what follows.

\section{UV luminosity function of high redshift galaxies}

The luminosity function and SFR density as a function of $z$ in our models
will depend on the parameters associated with star formation, reionization
in addition to the standard cosmological parameters. 
The parameters related to star formation like the IMF,
metallicity and $f_*$  could depend on redshift.  Since the exact evolution
of most of these quantities are difficult to predict, we compute
luminosity function at a given $z$ for a range of these parameters.

\subsection{Observed luminosity functions}

The observationally determined luminosity functions are taken from 
Sawicki et al. (2006) for $z=3$ and $z=4$, Iwata et al. (2003; 2007) for $z=5$  and
Bouwens \& Illingworth (2006) for $z=6$. Our model luminosity function is
computed using luminosity at $\lambda=1500~$\AA. We obtain the observed
luminosity function at $\lambda=1500~$\AA~assuming flat
spectrum in $f_\nu$.  
For $z>6$ we have constraints from three sets of observations. One is the upper limit
based on the tentative detection of three candidate galaxies at $8<z<12$ by Bouwens et al. (2005) 
in the HUDF. The limiting magnitude of the Bouwens et al. (2005) survey varies 
from field to field, ranging from apparent magnitudes in the AB system of $27.2$ 
to $28.7$. The three candidate galaxies have H$~\simeq 28$ and 
are detected in the two deep parallel fields 
observed with NICMOS covering a total area of 2.6 arc min$^2$ with $5\sigma$
limiting magnitudes of $28.5$. The second observational constraint comes from 
the upper limits on the luminosity function at $z = 7$ derived by
Mannucci et al. (2007) based on the absence of $z_{850}'$ galaxies.
The third study of relevance to high redshift star 
forming galaxies, is the deep near-IR imaging in the field of lensing clusters
by Richard et al. (2006). In this study, Richard et al. (2006) derive
the average luminosity function of galaxies at $6\lesssim z \lesssim 10$.

\subsection{Modeling the luminosity functions at $3\leq z \leq 6$}
\begin{figure}
\centerline{
\psfig{figure=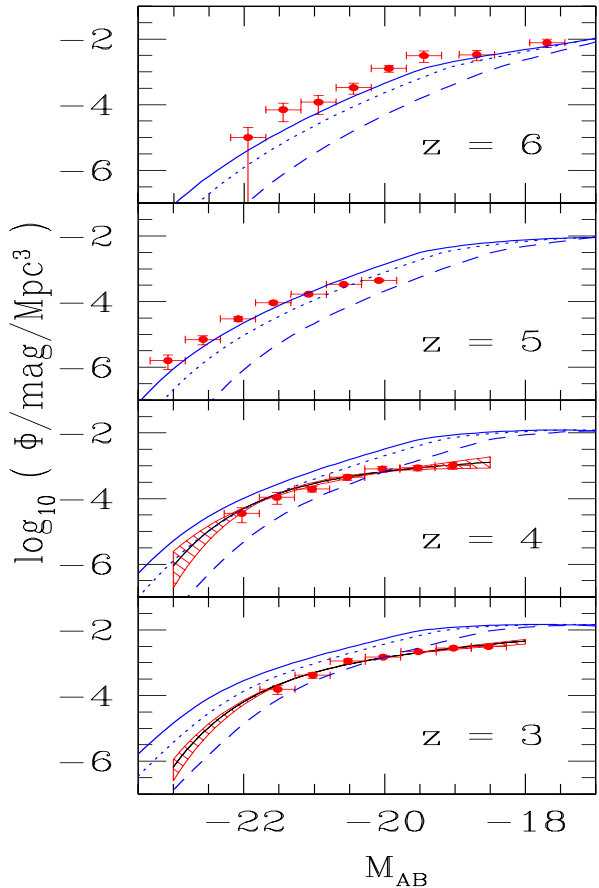,height=12cm,width=8cm,angle=0.0,bbllx=19bp,
bblly=144bp,bburx=286bp,bbury=695bp,clip=true}}
\caption[]{ Luminosity function at different redshift bins.
The solid, dotted and dashed curves are the predictions of models
with lower mass cutoffs $1$, $0.5$ and $0.1$ $M_\odot$ respectively.
All the models assume Salpeter IMF with upper mass cutoff $100~M_\odot$,
metallicity $Z=0.0004$, $\kappa = 1.0$ and $f_*=0.5$.
}
\label{figlf1}
\end{figure}

In this section, we present theoretically computed luminosity functions at $3\leq z\leq 6$. 
These are calculated using the formalism described in section~\ref{semianalytics}.
As reionization occurs at $z_{re}>6$ the luminosity functions in this redshift range are
not sensitive to the details of reionization history.
In Fig.~\ref{figlf1}, we overplot our computed luminosity functions on the observed 
luminosity functions at different redshift bins. The curves are for a Salpeter IMF 
with an upper mass cutoff $m_{\rm up}= 100~M_\odot$ and different lower mass 
cutoffs of $m_{\rm low} = 1,~ 0.5$ and $0.1~M_\odot$, adopting
a metallicity $Z=4\times10^{-4}$ (i.e. $0.02 ~Z_\odot$), 
$\kappa=1$, and $f_*=0.5$.  We find that for the IMF considered above
in a continuously star forming region the luminosity at $1500~$\AA~depends
very weakly on metallicity. For example changing $Z$ from $10^{-7}$ to 
$4\times 10^{-3}$ produces a maximum change of $0.17$ dex (Leitherer et al. 1999; Schaerer 2003). 
Thus we do not vary the metallicity in our models. 
The amount of reddening corrections we need
to apply is an unknown quantity. In principle this can depend on the redshift. 
To start with we have applied an uniform reddening correction 
by a factor $\eta = 4.5$ (found by Reddy et al. (2006) at $z\simeq 2$) at all redshifts.

It is very encouraging to see our semi-analytic model with a simple 
prescription for continuous star formation reproduces the observed 
luminosity function reasonably well over the redshift range of 
interest here. 
Also as described before, these models have $\tau_e$  consistent with the WMAP 3rd year data 
(see Table~\ref{tab_reion}).
The flattening in the predicted luminosity function seen
at the low luminosity end is due to the photoionization feedback
we apply to the halos with $v_c\le 90$ km s$^{-1}$.
It is clear from the figure that the observed luminosity function
at $z=5$ and $6$ are well reproduced by our models with $m_{\rm low} = 1~ M_\odot$.
The models with $m_{\rm low} = 0.1 ~M_\odot$ under predict the luminosity
functions at the high luminosity end by more than an order of magnitude.
Basically, lowering $m_{\rm low} $ from $1$ to $0.1 ~M_\odot$ makes the individual halos 
with a given star formation rate to appear $\sim 1$ mag fainter and moves the
luminosity function along the x-axis towards the low luminosity (high AB mag) end.
Thus in order to explain the observed luminosity function with
$m_{\rm low}<1~M_\odot$, $f_*/\eta$ has to be higher than $0.5/4.5$.

However, our model with  $m_{\rm low} = 1~ M_\odot$ 
over produces the luminosity function by more than $0.4$ dex at a given luminosity at 
$z=3$ and $4$ (Fig.~\ref{figlf1}).
It is clear from the figure that our models can reproduce the brighter end ($M_{\rm AB}<-20$) of the 
luminosity function for $0.1\le m_{\rm low}(M_\odot)\le 0.5$ at these redshifts. 
Thus it appears that a good agreement can be obtained by decreasing
$m_{\rm low}$ with decreasing redshift keeping $f_*$ and $\eta$ constant.
Such an evolution of $m_{\rm low}$  
may naturally be obtained due to
the increasing enrichment of the gas with the metals, 
by the previous generation of stars, as one goes to lower redshifts.
At the same time, decreasing $f_*$ (or increasing $\eta$) with decreasing
redshift keeping the IMF constant will also provide similar fits. 

\begin{figure}
\centerline{
\psfig{figure=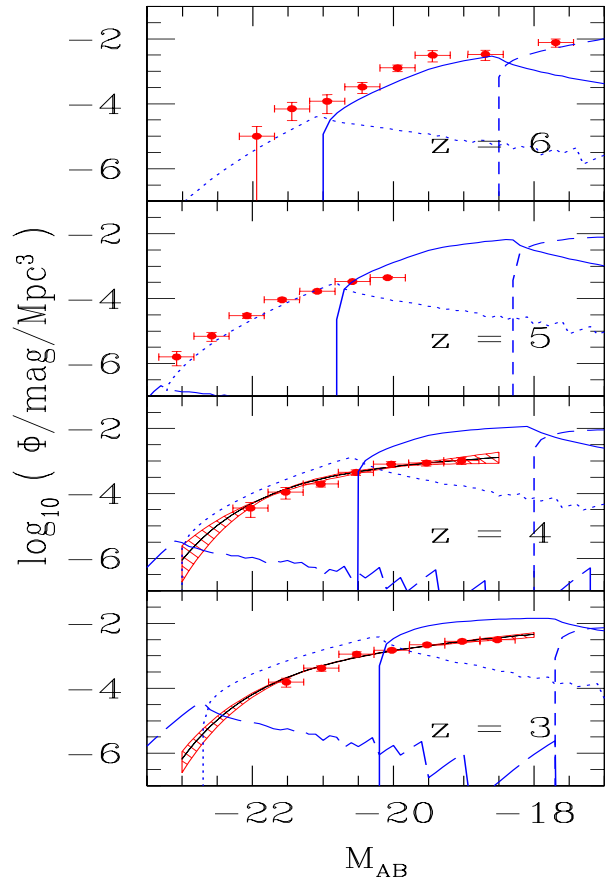,height=12cm,width=8cm,angle=0.,bbllx=19bp,
bblly=138bp,bburx=286bp,bbury=695bp,clip=true}}
\caption[]{ Contribution of halos with different mass ranges to the luminosity function.
Curves are shown for mass range $10^9-10^{10}~M_\odot$ (short-dashed), $10^{10}-10^{11}~M_\odot$ (solid),
$10^{11}-10^{12}~M_\odot$ (dotted) and $>10^{12}~M_\odot$ (long-dashed). The model presented here assumes
Salpeter IMF with lower mass cutoff $1~M_\odot$ and upper mass cutoff $100~M_\odot$,
$Z=0.0004$, $\kappa = 1.0$ and $f_*=0.5$.
}
\label{figlfrange}
\end{figure}

Even though our models broadly reproduce the observed luminosity functions,
it is obvious from Fig.~\ref{figlf1} that they over-produce the number of 
objects at lower luminosities especially at lower redshifts. Note that the photoionization 
feedback we use affects star formation in halos with $v_c<90$ km s$^{-1}$. This is marked by 
the break seen in our model luminosity functions at lower luminosity. Clearly this 
break occurs at absolute magnitudes
$M_{\rm AB}>-19.5$ suggesting that some additional feedback may be 
needed to suppress star formation even in halos with $v_c$ higher than $90$ km s$^{-1}$.
To know the range of mass where we still need more suppression of star formation 
we have plotted the contribution of different mass range to the luminosity function in 
Fig.~\ref{figlfrange}.
It is evident from this figure that one needs
more suppression in halos with mass $10^{10}-10^{11}~M_\odot$.  
This will be the rough range even if we consider  $m_{\rm low} = 0.1~M_\odot$.
In our models  $10^{11}~M_\odot$ corresponds to $v_c = 130$ and $145$ km s$^{-1}$
respectively for $z = 3$ and $4$.
Perhaps starburst driven galactic winds from these halos may provide such a 
negative feedback.
Pettini et al. (2001) have reported large scale outflows from $z\sim 3$
Lyman break galaxies that have a typical dynamical mass of about $10^{10}
~M_\odot$.
Recently, Croton et al. (2006) have noted that
the expulsion of hot gas due to supernova feedback affects 
halos with $v_c$ up to $200$~km~s$^{-1}$ in their semi-analytic models
implemented on the millennium dark matter simulations.
Thus while radiative feedback alone gives the correct shape of 
the observed luminosity function for $z\ge 5$, we need additional feedback for halos
$95\lesssim v_c({\rm km~s^{-1}}) \lesssim 150$ in order to explain the $z\sim 3$
luminosity functions.
Further, it is clear from Fig.~\ref{figlfrange} that luminosity
functions over the observed range, are insensitive to
the exact nature of the high mass cutoff in Eq.~(\ref{eqnsup}), as long as the 
star formation is suppressed in massive halos with 
$M>10^{12}~M_\odot$.

There are other parameters in our model that will change the 
luminosity of a given halo.
We explore the effect of these model parameters on the luminosity function
below. For reference,
we will take the model with $m_{\rm low} = 1~M_\odot$, and $Z=0.0004$
and other parameters as above as the fiducial model (called Model A).
Note that the reionization history for this model is also denoted as
model A in Table~\ref{tab_reion}.
\begin{figure}
\centerline{
\psfig{figure=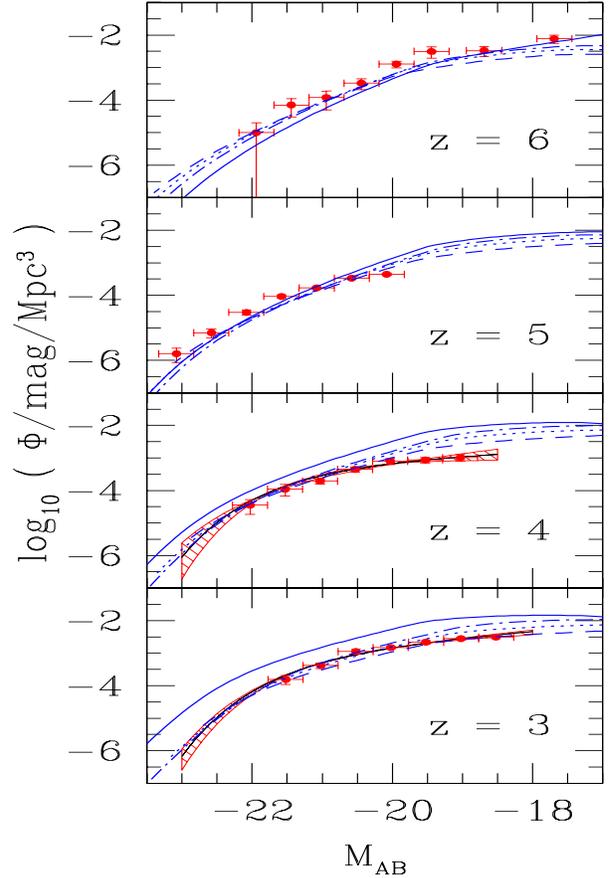,height=12cm,width=8cm,angle=0.,bbllx=17bp,
bblly=144bp,bburx=285bp,bbury=700bp,clip=true}}
\caption[]{ Effect of $\kappa$ on the predicted luminosity function.
Curves are shown for $\kappa = 1.0$ (solid),
$0.50$ (dot-dashed), $0.20$ (dotted) and $0.11$ (dashed).
All of them are for Salpeter IMF with lower mass cutoff $1~M_\odot$
and upper mass cutoff $100~M_\odot$ and metallicity $0.0004$.
For  $\kappa = 1.0$ we have assumed $f_*=0.5$. For the lower values of $\kappa$
we have used the lower values of $f_*$ to match the observations. 
}
\label{figlf3}
\end{figure}

First, we study the sensitivity of our results to the mode of star formation.
The burst mode of star formation corresponds to the limit $\kappa \to 0$.
We show in Fig.~\ref{figlf3}, the luminosity function for various redshifts taking different
values of $\kappa = 1/9, 1/5, 1/2, 1$. The solid curves in the figure are for 
our fiducial model A. For models with $\kappa<1$, we suitably scale $f_*$ to match the
observed luminosity function. 
As one decreases $\kappa$ and
goes towards the burst mode of star formation,
the number of objects in the brighter end of the luminosity function,
significantly increases. However better matching to the data can be 
obtained by lowering the value of $f_*$. For example, at $z=6$ models
reproduced the observed luminosity function for $f_* = 0.32,~ 0.20$ and
$0.13$ for $\kappa = 1/2,~ 1/5$ and $1/9$ respectively. Similarly at $z=3$ we need
$f_* = 0.17,~ 0.10$ and $0.07$ for $\kappa = 1/2,~ 1/5$ and $1/9$ respectively.
However, in order to keep the $z_{re}$ high enough we need to preserve $f_* f_{esc}$
by increasing $f_{esc}$
whenever we decrease $f_*$ from $0.5$.
Therefore, in the framework of models discussed here observations of 
luminosity function at $z\lesssim 6$ can be reproduced by both the 
continuous as well as burst mode of star formation. 
However, in both the cases we need to allow for redshift evolution of
either of $m_{\rm low}$, $\eta$ or $f_*$.  Decrease in $m_{\rm low}$ and increase
in $\eta$ with decreasing redshift can naturally arise with the
expected redshift evolution of metallicity. 

The nature of the IMF and duration of the star formation activities in a given galaxy
can be obtained by fitting the observed spectral energy distribution (SED) with synthetic 
spectrum. Eyles et al. (2007) have fitted the rest frame UV-optical SED of $30$ galaxies
at $z \sim 6$ with reliable photometric or spectroscopic redshifts. They found a
surprisingly large fraction of galaxies with a signatures of substantial 
Balmer breaks indicating the
presence of an underlying old stellar population that dominates the stellar masses.  The
calculated age of these objects are in the range 180-640 Myr and stellar masses in
the range $1-3\times10^{10}~M_\odot$.  It is interesting to note that in our 
model A that fits the $z=6$ luminosity function reasonably well, the observed 
range in the luminosity is produced by halos with stellar masses in range 
$3\times10^8-2\times10^{10}~ M_\odot$. The dynamical time-scale at this
epoch is 124 Myr and thus the expected time-scales for the star formation 
activity in these models, are consistent with that noted by Eyles et al. (2007). 
When we consider an IMF with $m_{\rm low} = 1~M_\odot$ the Balmer break naturally 
occurs in the spectrum. No break will be visible in the photometric data 
if one uses $m_{\rm low} = 10~M_\odot$. Thus our models which fit the luminosity
function will also be consistent with the SED of few of the galaxies observed
by Eyles et al. (2007).
When we consider $\kappa=1$, we require roughly $50\%$ of the baryon mass to
go through star formation over few dynamical time-scale. This is again
consistent with the median gas fraction of $50\%$ and the corresponding
stellar mass inferred from the high-$z$ Lyman break galaxies (Erb et al. 2006).

\begin{figure}
\centerline{
\psfig{figure=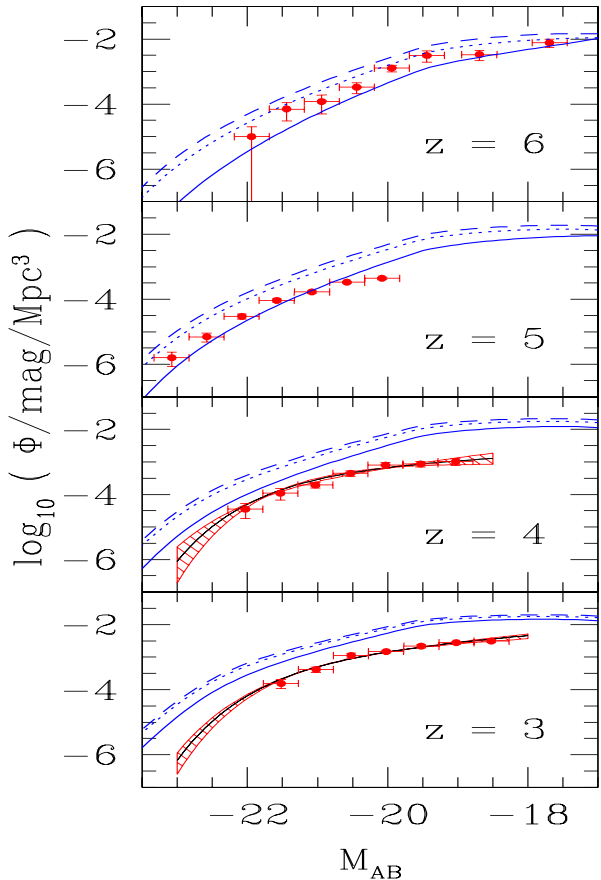,height=12cm,width=8cm,angle=0.,bbllx=17bp,
bblly=144bp,bburx=285bp,bbury=700bp,clip=true}}
\caption[]{Dependence of luminosity functions on $\sigma_8 $ and $n_s$.
Solid curves are for $\sigma_8 = 0.75$
and $n_s=0.95$ i.e. the WMAP 3rd yr cosmological parameters. The dotted lines are for
$\sigma_8 =0.85$ and $n_s = 0.95$. The dashed lines represent the WMAP first yr
parameters i.e. $\sigma_8 =0.85$ and $n_s = 1.00$.
All of them are for Salpeter IMF with lower mass cutoff $1~M_\odot$
and upper mass cutoff $100~M_\odot$ and metallicity $0.0004$.
We have taken $\kappa = 1.0$ and $f_*=0.5$.
}
\label{figlf4}
\end{figure}

Finally, in Fig.~\ref{figlf4}, we have considered the sensitivity of 
our results to changes in $\sigma_8$ and $n_s$, from the values favored by 
WMAP 3yr data. As expected an increase in $\sigma_8$ or $n_s$ leads to larger
number of objects at any given luminosity, at all redshifts.
This basically reflects the fact that increasing $\sigma_8$ or
$n_s$ increases the number of collapsed halos. So the same data
requires a lower value of $f_*$, for higher $\sigma_8$ or $n_s$.

We see therefore that combining a fairly simple model of
star formation with the modified Press-Schechter formalism 
(according to the Sasaki prescription) one 
can fit the whole range of high redshift 
observed galaxy luminosity functions from $z=3$ to $z=6$, for a reasonable range
of parameters. The feedback due to photoionization is sufficient to explain the 
$z=6$ luminosity function. Whereas we need additional feedback, possibly due
to galactic scale super winds driven by supernovae, to explain 
the low luminosity end for $z=3$.
We now proceed to our model predictions for the higher redshift range.

\section{Constraining star formation at \lowercase {$z>6$}}

In this section, we compare  our model predictions with observations at $z >6$.  
As one expects the epoch of reionization to fall in this redshift range (see 
Tables~\ref{tab_reion} and \ref{tab_reion_m}),  our model predictions for a 
given $z$ will be  very sensitive 
to the reionization history. In addition one may need to consider the effects of 
molecular cooled halos prior to reionization. At redshifts $z>6$, in the absence of 
spectroscopic redshift measurements we have observational constraints in the 
form of integrated source counts and average luminosity functions obtained
over a large redshift range. Here, we make predictions for both sets of 
observations, so as to probe the nature of star formation at such high redshifts. 

\subsection{UV Luminosity function}

\begin{figure}
\centerline{
\psfig{figure=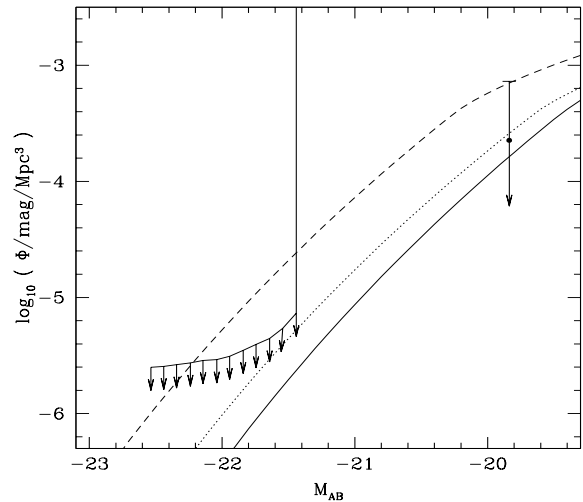,height=7cm,width=8cm,angle=0.0}}
\caption[]{Luminosity function at $z=7$. The observed data points are taken
from Mannucci et al. (2007). The upper limit at M$_{AB}= 19.8$ is from Bouwens
et al. (2004). The solid and dashed curves are for
models A and F as in Table~\ref{tab_reion} respectively. The dotted
line is for the models with $\kappa = 0.5$, $f_*$ = 0.32 and rest
of the parameters as in model A. 	
}
\label{figlfz7}
\end{figure}

First we consider the upper limits on the luminosity function at $z=7$ given 
by Mannucci et al. (2007). In Fig.~\ref{figlfz7} we show the observed 
luminosity function as well as the theoretically predicted luminosity functions at
$z=7$. The continuous curve is for model A and dotted curve in Fig.~\ref{figlfz7} is
for model A with $f_* = 0.32$ and $\kappa = 0.5$. Both these models fit the 
$z =6$ luminosity function well (see section~4). It is clear from the figure that the 
luminosity function predicted by these models are consistent with the null detection 
of galaxies by Mannucci et al. (2007). Note we have used $\eta = 4.5$
in our calculations. From the figure we can infer that a slightly lower value of $\eta$
is also allowed by the observations. If we just follow the line of arguments
we have presented in the last section, for $z\ge 6$ we expect the $m_{\rm low}$ 
to be higher than $1~M_\odot$. The dashed line is for model F that has  $m_{\rm low} =10~
M_\odot$ and $\eta=4.5$. Clearly this model over produce the 
number of high luminosity objects. The difference will become wider if we use a lower
value of $\eta$. Thus, the upper limits in the luminosity 
function at $z=7$ can be understood as due to just the
effect of redshift evolution 
of dark matter halos from the standard structure formation, without a strong
evolution in the nature of star formation.
Labbe et al. (2006) using the Spitzer observations of candidate
galaxies at $z\sim 7$ in UDF found that these
galaxies have typical stellar mass of $1-10\times 10^{9}~M_\odot$
by fitting the SED.
The typical age of these galaxies are $50-200$ Myr with average star
formation rate of $25~M_\odot$ yr$^{-1}$ assuming a constant star formation
model. The age of these galaxies are then consistent with $\kappa \gtrsim 1/2$
in our model. This confirms that the detected candidates are 
undergoing prolonged star formation activities consistent with our model prediction.

The observational situation, however is not that simple
as Richard et al. (2006) have reported the detection of a large number of $z\ge6$ 
candidate galaxies in their search towards strong lensing clusters. 
%
\begin{figure}
\centerline{
\psfig{figure=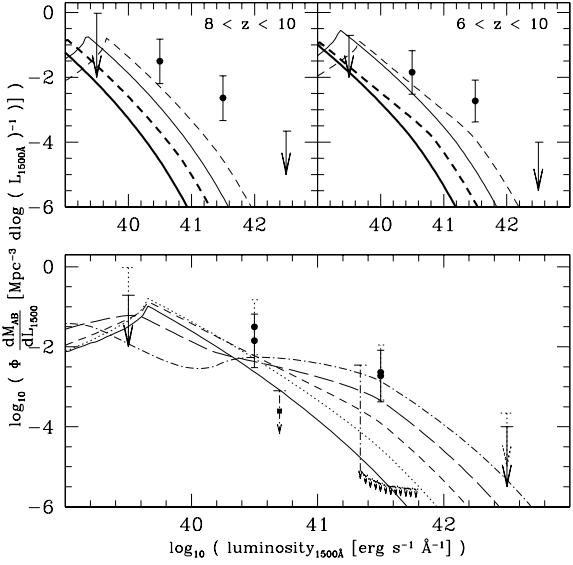,width=8.0cm,angle=-0.0
}}
\caption[]{Luminosity function for $z=9$ (top left panel) and $z=8$
(top right panel). 
The solid and dashed lines corresponds to model A and model F respectively.
The thick lines assume $\eta=4.5$
while the thin lines are for $\eta=1$. 
Bottom panel shows the redshift evolution of luminosity function for
model F with $\eta=1$. Curves are drawn for
$z=10$ (solid), $9$ (dotted), $8$ (short-dashed), $7$ (long-dashed) and
$6$ (dash-dotted). The observed points for $ 6<z<10$ (range 2) and $8<z<10$
(range 1) are shown with errorbar in solid and dashed lines respectively. The
Luminosity function at $z=7$ from Mannucci et al. 2007 are given by dot-dashed
lines.
}
\label{fig_richard1}
\end{figure}
%
\begin{figure}
\centerline{
\psfig{figure=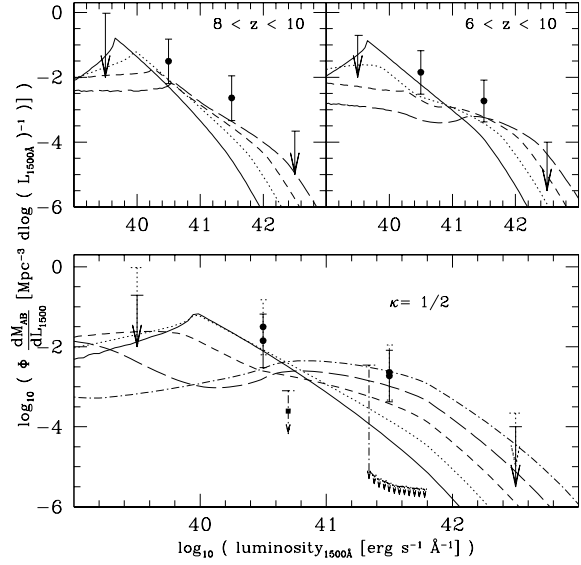,width=8.0cm,angle=-0.0}}
\caption[]{The variation of the luminosity functions for $z=9$ (top left panel) 
and $z=8$ (top right panel) with $\kappa$. All models assume the top heavy IMF,
$\eta=1$ and other parameters as in model A.
The curves are for $\kappa=1$ (solid), $\kappa=1/2$ (dotted), $\kappa=1/4$ (short-dashed)
and $\kappa=1/9$ (long-dashed).
Bottom panel shows the redshift evolution of luminosity function for
$\kappa=1/2$. Curves are drawn for
$z=10$ (solid), $9$ (dotted), $8$ (short-dashed), $7$ (long-dashed) and
$6$ (dash-dotted). The observed points for $ 6<z<10$ (range 2) and $8<z<10$ (range 1) 
are shown with errorbar in solid and dashed lines respectively. The
Luminosity function at $z=7$ from Mannucci et al. 2007 are given by dot-dashed
lines.
}
\label{fig_richard2}
\end{figure}
In Fig.~\ref{fig_richard1} we compare our model predictions with that obtained 
by Richard et al. (2006). It is important to note that
spectroscopic redshifts are not available for the objects detected
by Richard et al. (2006). Thus they have obtained only the average luminosity function 
either in the redshift range $8<z<10$ (called range 1) or $6<z<10$ (called range 2).
Our model predictions are computed at the redshifts in middle of these ranges.
In the top left and right panels in Fig.~\ref{fig_richard1}, we have shown our model 
predictions for $z=9$ (for range 1) and $z=8$ (for range 2) respectively. 
The thick solid and dashed line are {respectively} for model A and the top-heavy 
model F with $\eta = 4.5$.
The corresponding thin lines are for $\eta = 1$. Clearly the luminosity functions
derived from Richard et al.'s data can not be explained by simply changing $m_{\rm low}$
as we have done for $z \le 7$.
Even the model with no extinction correction and a top heavy IMF (i.e. $m_{\rm low} = 
10~ M_{\odot}$) under predicts the abundance of the higher luminosity galaxies 
inferred by Richard et al. Our predictions for the luminosity function in range 2 is
in slightly better agreement with the data when we use $\eta = 1$ than that for 
the range 1.

To investigate this issue further, in the bottom panel of Fig.~\ref{fig_richard1} 
we show the redshift evolution of luminosity function for the model with no extinction 
correction ($\eta=1$) and a top heavy IMF. The lines are for $z=10$ (solid), $9$ (dotted), 
$8$ (short-dashed), $7$ (long-dashed) and $6$ (dash-dotted). The observed 
luminosity functions for range 1 and range 2 are shown with errorbar in dashed 
and solid lines respectively. 
Also the upper limits on the observed luminosity function at $z = 7$ are shown with arrows.
The sharp cutoff seen in the low luminosity end of the 
luminosity function for $z\ge 8$ is due to the cooling cutoff
at $T_{\rm vir}=10^4$ K. The flattening in the luminosity function for $z\le7$ seen in the 
figure is due to radiative feedback, as reionization in this model occurs 
at $z_{re}=8.4$ (see Table~\ref{tab_reion} for details).
All our models clearly under produce the luminosity function at $8<z<10$. 
It is also clear from the figure that this model also over produces the 
abundance of 
$z=7$ galaxies compared to that inferred by Mannucci et al. (2007)
(See Fig.~\ref{fig_richard1}).
Thus to fit the Richard et al.'s data one has to increase the luminosity 
of individual galaxies only at $z\ge8$. 

Next we investigate whether going over to a burst mode of star formation will yield
a better match to Richard et al.'s data. 
In order to examine this possibility, we have shown in Fig.~\ref{fig_richard2} 
our model predictions for the luminosity functions at $z=9$ (right panel) and $z=8$ (left panel) 
for a range of $\kappa$. The models also assume a top heavy IMF with 
a mass range $10-100~M_\odot$, no extinction correction (i.e. $\eta = 1$) 
and all the other parameters as in model A.
The curves are for $\kappa=1$ (solid), $\kappa=1/2$ (dotted), $\kappa=1/4$ (short-dashed)
and $\kappa=1/9$ (long-dashed). 
We see that a moderate decrease
in the value of $\kappa$ by a factor of $2-4$ could make
the model consistent with the Richard et al. data, especially if these
galaxies are at $z \sim 8$ (also see bottom panel of 
Fig.~\ref{fig_richard2} for the redshift evolution of the luminosity
function for model with $\kappa = 1/2$).
A decrease of $\kappa$ to even smaller values, say to $\kappa =1/9$
leads to a decrease in the number of lower luminosity galaxies,
at $z=8$, but 
matches the Richard et al. data if the detected
galaxies had a redshift range $8<z<10$.
Even this agreement is only with the lower end of 
the numbers allowed by the error bars given by Richard et al.

\begin{figure}
\centerline{
\psfig{figure=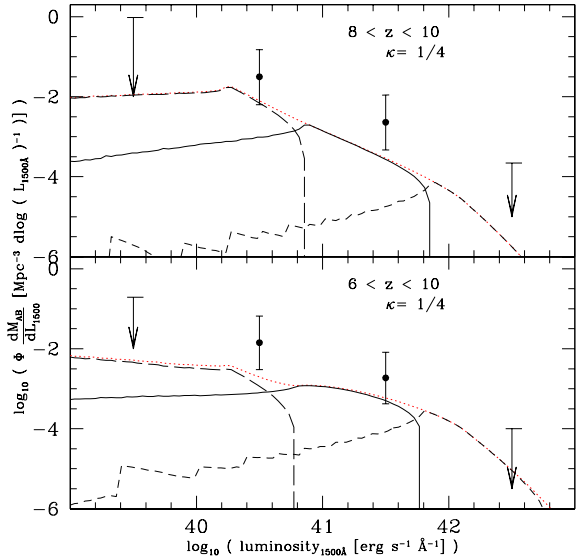,width=8.0cm,angle=00.0}}
\caption[]{Contribution to the luminosity function by different mass range. Curves
are drawn for a top-heavy IMF ( $10-100~M_\odot$ ), $\kappa = 1/4$ and $\eta = 1.0$
for $z=9$ (top) and $z=8$ (bottom). Long dashed lines are for contribution coming
from halos with mass $M\le 10^{9}~M_\odot$, solid lines are for $10^9~M_\odot\le M \le
10^{10}~M_\odot $ and short dashed lines are for mass $M\ge 10^{10}~M_\odot$. Dotted lines
represent the total luminosity functions.
}
\label{mrange}
\end{figure}

In Fig.~\ref{mrange} we plot the mass range contributing to different luminosity
ranges for $\kappa = 1/4$ for $z=8$ and $z=9$. The luminosity of galaxies
that are detected by Richard et al.(2006) are produced by galaxies with dark matter
mass of $\gtrsim 10^9~ M_\odot$, and for our models this corresponds to
a stellar mass of $\gtrsim 10^8~ M_\odot$. This is roughly consistent with
the stellar mass estimation of Richard et al. (2006). The objects with luminosity
greater than $10^{42}$ erg s$^{-1}$\AA$^{-1}$~are produced by dark matter halos with 
$M\ge 10^{10}~M_\odot$. Our model predictions are consistent with the 
absence of such very bright galaxies. Richard et al. (2006) obtained a SFR of
individual galaxies in the range $10-40~ M_\odot$ yr$^{-1}$ using template fitting 
method. In our case the SFR of the galaxy is a function of time typically
lasting for $4\kappa$ times the dynamical time-scale. However,
we can write the average SFR in a given halo as,
\begin{equation}
SFR = \left({10\over \kappa}\right) \left({M\over 10^9 M_\odot}\right) \left({f_*\over 0.5}\right) 
\left({1+z\over 10}\right)^{3/2} ~M_\odot~yr^{-1}
\end{equation}
For $\kappa=1/4$ the typical average SFR in the halos in our model is $40~ M_\odot$~yr$^{-1}$.
This is consistent with the range found by Richard et al. (2006) (see their Tables C2 and C3). 
%
\begin{figure}
\centerline{
\psfig{figure=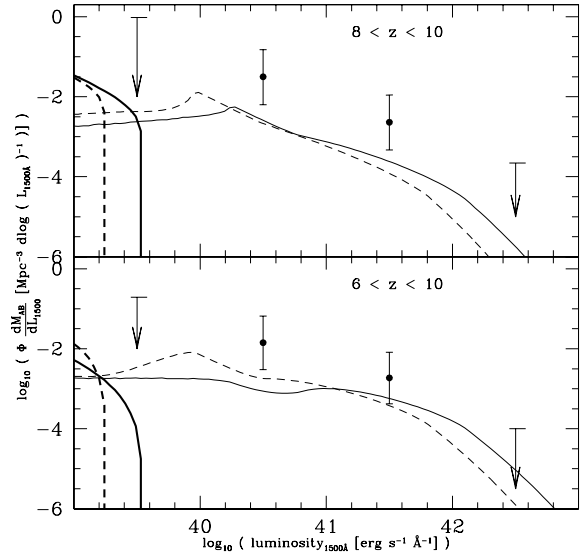,width=8.0cm,angle=00.0}}
\caption[]{Luminosity function calculated for $z = 9$ and $10$ when we add the
contribution from the molecular cooled halos. We have assumed a top heavy
IMF from $10$ to $100 ~M_\odot$ with $\eta = 1.0$. For atomic cooled halos
$f_* = 0.50$ and for molecular cooled halos $f_* = 0.10$. 
We have shown the luminosity
functions for $z=9$ (top) and $z=8$ (bottom) with $\kappa =
1/4$ (solid line)  and $1/2$ (dashed line). The contribution coming from
the molecular cooled halos are shown with thick lines where as thin lines
represent the contribution from atomic cooled halos.
}
\label{fig_lf_richard_molecular}
\end{figure}

In all the models discussed till now the effect of molecular cooled halos 
has not been considered.
We expect star formation to be going on in some
of the molecular cooled halos at least prior to the epoch of reionization
and such a star formation is thought to be a
very important component of contemporary models of reionization.  
Fig.~\ref{fig_lf_richard_molecular} shows the expected range in luminosity from the 
molecular cooled halos with $f_* = 0.1$ (thick lines). We have drawn curves for top-heavy 
IMF (i.e. $10-100~M_\odot$),
$\eta = 1.0$ and $\kappa = 1/4$ (solid) and $1/2$ (dashed).
Clearly the expected luminosity is below the detection limits achieved in 
the present day deep imaging surveys. Thus even if the molecular cooled halos
are present in the early universe they will not contribute to the
observed luminosity function (or for that matter to the inferred star formation rate density).
However, star formation in molecular cooled
halos provides additional Lyman continuum photons there by making 
reionization to occur earlier (see Table~\ref{tab_reion_m}). 
This implies a relatively greater suppression of low mass halos due
to reionization feedback, and in turn will lead
to more and more difficulty in explaining the Richard et al. points. 
Note that while plotting Fig.~\ref{fig_lf_richard_molecular} we have not considered
the contribution of molecular cooled halos to the reionization in a self-consistent
way. We will discuss the effect of reionization feedback on the 
predicted luminosity function in section~6.

\subsection{Integrated source count at $8\le z\le12$}

\begin{figure}
\centerline{
\psfig{figure=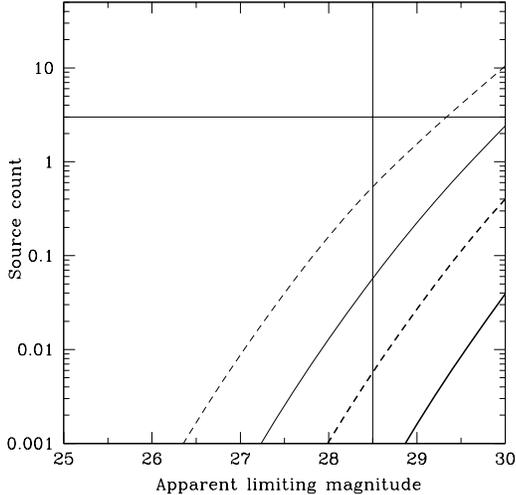,height=7.0cm,width=7.0cm,angle=-0.0}}
\caption[]{Integrated galaxy count as a function of apparent magnitude
for the $8<z<12$ in an $2.6$ arcmin$^2$ survey area.
We have considered parameters as in model A, except for varying $\eta$ and
$m_{\rm low}$. The thick and thin solid lines show the source count predictions for
model A with $\eta=4.5$ and $\eta=1$ respectively for $m_{\rm low} = 1 M_\odot$.
The thick and thin dashed lines shows the
corresponding source counts for a top heavy mass function
with $m_{\rm low} = 10 M_\odot$, and all other parameters as in model A.
The horizontal line corresponds to a detection of three galaxies in the
above redshift range. The vertical line shows the $5\sigma$ limit of
$28.5$, for the Bouwens et al. detections.
}
\label{figsc}
\end{figure}
\begin{figure}
\centerline{
\psfig{figure=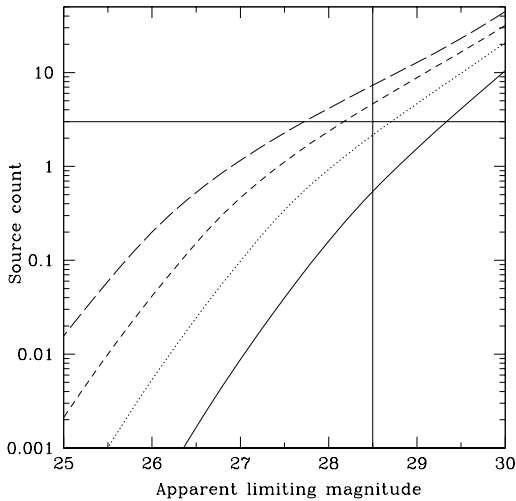,width=7.0cm,angle=-0.0}}
\caption[]{
Total number of galaxy count as a function of apparent magnitude
for the $8<z<12$ in an $2.6$ arcmin$^2$ survey area,
for the models of Fig.~\ref{fig_richard2} (top panels). The line styles are
are the same as in Fig.~\ref{fig_richard2}.
The horizontal line corresponds to a detection of three galaxies in the
above redshift range. The vertical line shows the $5\sigma$ limit of
$28.5$, for the Bouwens et al. detections.
}
\label{figsc2}
\end{figure}

In Fig.~\ref{figsc}, we show our model predictions of integrated source 
count at $8<z<12$ as a function of the limiting apparent magnitude 
for an area of $2.6$ arcmin$^2$ (as in Bouwens et al. 2005). The redshift
range covered is similar to the range 1 in Richards et al. (2006).
The thick and thin solid lines show the predictions for model A 
with $\eta=4.5$ and $\eta=1$ respectively. The corresponding dashed lines
are for $m_{\rm low} = 10~M_{\odot}$. From Fig.~\ref{figsc}, it is clear that
the integrated source counts predicted by the continuous star formation 
models (with $m_{\rm low}\le 10 ~M_\odot$) will always be below the 
upper limit obtained by Bouwens et al. (2005) irrespective of $f_*$ and $\eta$. 
However, if the 3 candidate galaxies tentatively identified in the 
Bouwens et al. (2005) data become confirmed as high-$z$ galaxies, then our continuous 
star formation models will fail to reproduce their abundance. 

We have also computed the source counts for the burst models 
discussed in Fig.~\ref{fig_richard2}. They are shown in Fig~\ref{figsc2}.
The models with a top heavy IMF and
with $\kappa > 1/2$ predicts number counts 
less than the upper limit given by Bouwens et al., while
the model with $\kappa=1/4$ gives counts slightly larger.
However the model with $\kappa=1/9$ over predicts the 
number counts and so is probably ruled out by the
Bouwens et al. data.
It appears that having a top heavy IMF and a moderate decrease
of $\kappa$ can make our models consistent with both 
the Richard et al. and the Bouwens et al. data if one allows for $2\sigma$ errors.
Note that one of the uncertain factors could be the effect of amplification
bias in the estimation of luminosity function by Richard et al. (2006).

In summary, we have reproduced different sets of observations available
in the literature for $z>6$. The upper limits on the luminosity function
for $z=7$ obtained by Mannucci et al. (2007) and the upper limits on
the integrated source counts for $8\le z\le 12$ as a function of limiting
apparent magnitudes by Bouwens et al. (2005) are consistent with the
continuous star formation models that fits $z\le 6$ luminosity function.
The decline in the number density of sources at high redshifts can just
be explained from the decline in the halo number density expected from 
structure formation models alone without any dramatic change in the
nature of star formation activities at $z\ge6$. 
In the language of observers these data at $z>6$ are consistent
with the pure number density evolution expected from the $\Lambda$CDM
model without any luminosity evolution.
However, these models
fail to reproduce the luminosity function inferred by Richard et al. (2006)
based on galaxies detected around strong lensing clusters. Such models
will also fail if the three candidate galaxies identified by Bouwens et al.
(2005) are confirmed as high-$z$ galaxies. For the 
cosmological parameters constrained by WMAP 3rd year data
both these observations can only be explained if star formation occurs in a 
burst mode (i.e. $\kappa < 1/2$) with high efficiency, top-heavy IMF and no reddening corrections 
for UV light. Thus if Richard et al. (2006) observations are correct then one 
needs a sudden change in the nature of the star formation  at $z\ge8$.
In other words we need a strong luminosity evolution on top of the
number density evolution at $z>8$.

Thus it is important to get a clearer picture from the observational side
before one can draw any firm conclusions on the nature of the star formation
at $z\ge6$. Nevertheless, the exercise presented here
clearly suggests that strong constraints on the nature of  
star formation can be obtained once there is improvement in the
observations. In particular accurately measured luminosity functions 
over small redshift intervals at $z> 6$ will provide important constraints
on the nature of reionization. We expand on this point in the following section.

\section{Probing the reionization history with \lowercase {$z\ge6$} luminosity function}

We now investigate further the possibility of using the redshift evolution of
the observed luminosity function to probe the reionization history of the universe.
In particular we are interested in the effect of molecular cooled halos and 
star burst activity during dark ages.
In our model calculations the luminosity function at the low luminosity end
is mainly produced by low mass halos that are
affected by radiative feedback.  
As discussed before, the radiative feedback results in a break in the luminosity
function that corresponds to halos with circular velocity of $90~$km s$^{-1}$
for $z<z_{re}$. However, the exact luminosity
at which this break will appear in the luminosity function depends on the 
amount ($f_*$) and duration ($\kappa~ t_{dyn}$) of star formation and the IMF
(for example the value of $m_{\rm low}$).
We illustrate to begin with, the effects of reionization feedback in a general manner, and then
more specifically in relation to Richard et al. data.

\begin{figure}
\centerline{
\psfig{figure=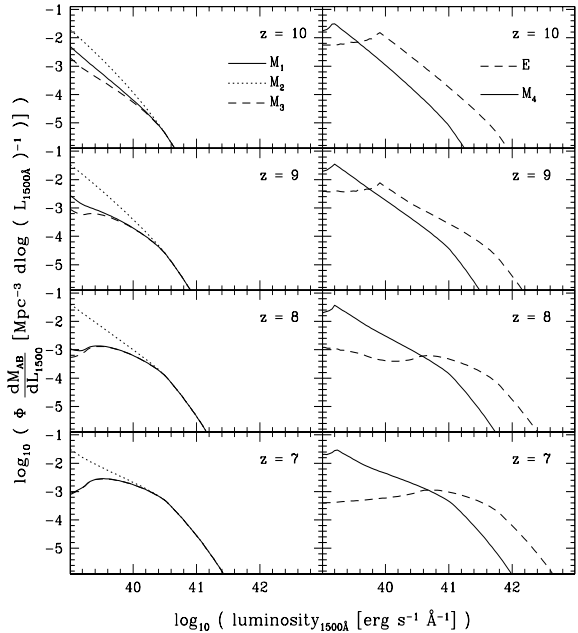,width=8.0cm,height=9.0cm,angle=0.}}
\caption[]{ Luminosity functions for $z=7,~8,~9$ and $10$ are plotted
from bottom to top. In the left side panels the solid, dotted and short-dashed 
curves are for our models M$_1$, M$_2$ and M$_3$ 
respectively (see Fig.~\ref{fig_reion} and Table~\ref{tab_reion_m} for reference).
All these models assume $\eta=4.5$.
The solid and long-dashed curves in the right side panels are for 
model E with $\eta = 1$  and model M$_4$ with $\eta = 4.5$ respectively.
}
\label{figlfzreion1}
\end{figure}

In the left side panels of Fig.~\ref{figlfzreion1}, we show the 
luminosity functions at different redshifts predicted by a set
of self-consistent models with molecular cooled halos incorporating reionization 
feedback. The solid, dotted and short-dashed
curves are for our models M$_1$, M$_2$, and  M$_3$ 
respectively, 
(see Fig.~\ref{fig_reion} and Table~\ref{tab_reion_m}).
Note that the model M$_1$ and M$_2$ differ only in $f_*$ for molecular cooled 
halos and both have atomic cooled halos as in model A.  Model M$_3$ is same as M$_2$ but
with $m_{\rm low}= 50$ M$_\odot$ in the molecular cooled halos.  
The redshifts of reionization in these models are $10.8$ , $5.9$  and $11.6$ respectively 
(see Table.~\ref{tab_reion_m}). It is clear from Fig.~\ref{figlfzreion1} that the luminosity 
function from these three models are identical at $L(1500)\gtrsim 10^{40.4}$ erg s$^{-1}$\AA$^{-1}$.
In the low luminosity end models M$_1$ and M$_3$ produces similar luminosity function
but differ significantly from that of model M$_2$.  
This difference is due to the early suppression of low mass halos in models M$_1$ and M$_3$ 
due to early reionization. It is clear from this illustration that even though the molecular
cooled halos are not directly detectable, their contribution to reionization can be probed
using an accurately measured redshift evolution of luminosity function at $z\ge6$.

In the right side panels of Fig.~\ref{figlfzreion1} we plot the results for model M$_4$
and E in solid and long-dashed lines respectively. In model M$_4$ we use $\kappa=1/4$ and
normal IMF in all the halos. Whereas in model E we use $\kappa = 1/9$
and top-heavy IMF. In these models star formation occurs over a
short time scale, compared to the models where $\kappa=1$.
As expected, both the models therefore
clearly produce a larger number of luminous galaxies at 
$L(1500)\ge10^{41.4}$~erg~s$^{-1}$\AA$^{-1}$. These two models have similar optical
depth to reionization, $\tau_e = 0.114$ for model E, and $\tau_e=0.121$ for
model M$_4$. However one can see from Fig.~\ref{figlfzreion1}, that
the slope of their luminosity functions, at the low luminosity
end, are very different.  The reason for this lies in the two models
having different $z_{re}$.
For model M$_4$, the redshift of reionization $z_{re}=6.2$.
As this is lower than the $z$ range over which we study the
luminosity function we do
not see a significant flattening in the luminosity function at low luminosities.
On the other hand, in the case of model E the reionization occurs at $z_{re}=8.4$.
The resulting sudden change in the luminosity
function at $z\le8$ compared to that at high-$z$
is clearly visible in Fig.~\ref{figlfzreion1}. The break in the luminosity occurs
around $L(1500)=10^{41}$~erg~s$^{-1}$\AA$^{-1}$. This example clearly demonstrates
that one can have detectable changes in the luminosity function close to  $z_{re}$.

In summary,
if the ionization feedback is the main contributor to the suppression of star formation 
activities in the low mass halos then the redshift of reionization can be constrained 
from the epoch at which the low luminosity flattening occurs in the galactic luminosity
function. In the presence of star bursting activities or low dust extinction we expect
the break luminosity to be occurring at $L(1500) \sim 10^{41}$ erg s$^{-1}$\AA$^{-1}$
that is easily detectable with the present day telescopes.
Hence, we can use the low end of the luminosity function at high-$z$
as an indicator to the reionization history whereas the high end
of the luminosity function can probe the mode of star formation.

\begin{figure}
\centerline{
\psfig{figure=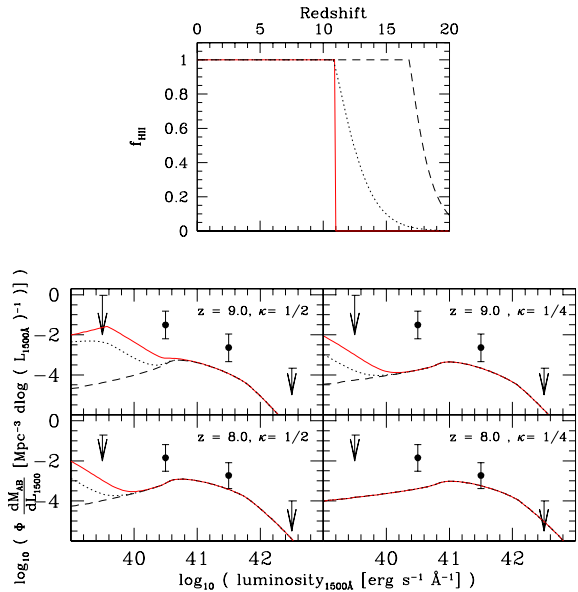,width=8.0cm,angle=0.}}
\caption[]{The effects of early reionization. Top panel
shows three fiducial reionization scenarios. The dashed line is for
$z_{re} = 16.8$ and $\tau_e = 0.236$, dotted line is for $z_{re} =
10.9$ and $\tau_e = 0.142$.
The solid line assume an abrupt reionization
at $z_{re} = 10.9$ which produces an optical depth of $\tau_e = 0.111$.
Rest four panels show the luminosity
functions obtained with these reionization models at different redshifts
for our top heavy model ($10 - 100 ~M_\odot$) with $f_* = 0.50$ and
$\eta = 1.0$. Middle panels show the the luminosity function
for $z = 9.0$ with $\kappa = 1/2$ (middle left panel) and $\kappa = 1/4$
(middle right panel). The bottom panels are for $z = 8.0$ with $\kappa =
1/2$ (bottom left panel) and $\kappa = 1/4$ (bottom right panel). We
follow the same line style as top panel for different reionization
models.
}
\label{figlfzreion}
\end{figure}

Now consider more specifically the constraints implied by the Richard et al. data.
As explained in the previous section to reproduce the observations of Richard et al. 
(2006) we need a burst mode of star formation with no reddening correction for 
UV light.
In all our
self-consistent atomic cooled models reionization occurs at $z_{re}< 10$
(see Table~\ref{tab_reion}). The effect of this is reflected in Figs.~\ref{fig_richard1} and 
\ref{fig_richard2} where the predicted number density of galaxies with luminosity of order 
$10^{40}$ erg s$^{-1}$\AA$^{-1}$~is higher at $z\ge 8$ than that at $z<8$
(whereas naively one would have expected it to be the other way round). 
Higher value of $f_{esc}$ or the inclusion of molecular cooled halos would produce 
reionization at higher redshifts (see Table~\ref{tab_reion_m}). We explore the 
effect of such early reionization to the predicted luminosity function for $z>6$ in 
Fig.~\ref{figlfzreion}.
%
In the top panel of Fig.~\ref{figlfzreion}, we present three fiducial reionization 
scenarios. The dashed line represents a very early reionization of $z_{re} = 16.8$ and the 
corresponding optical depth is $\tau_e = 0.236$.
This scenario is considered, for illustrative purposes,
in the view of 1st year WMAP data
(Spergel et al., 2003), (where a $\tau_e =0.17^{+0.08}_{-0.07}$ was favored). 
The dotted line represents a reionization model which has 
$z_{re} = 10.9$ and $\tau_e = 0.142$. In these two models the hydrogen ionization 
fraction changes in a continuous fashion. We have considered a third model (solid line) where 
reionization occurs abruptly at $z_{re}= 10.9$. This leads to an optical depth 
$\tau_e = 0.111$. This model mimics the reionization model taken by 
the WMAP team to get the reionization redshift from the electron optical depth 
using third year data (Spergel et al., 2006).
The rest of the panels of Fig.~\ref{figlfzreion} show the luminosity function obtained 
for above mentioned three reionization models. 
For the model with $z_{re} = 16.8$ (dashed curves) we see a clear
turnover in the luminosity function at $L(1500) \sim 10^{41}$ erg s$^{-1}$\AA$^{-1}$
due to photoionization suppression. Clearly it will become more difficult to explain
the luminosity function of Richard et al. (2006) especially for $L(1500)<10^{41}$
erg s$^{-1}$\AA$^{-1}$.
It is more interesting to note 
the presence of low luminosity galaxies in the other two cases. In these cases
the detected low luminosity galaxies are the ones that formed prior to
reionization. Number of such galaxies are larger when one considers a higher
value of $\kappa$ (or prolonged star formation activities). The difference between
the dotted and the dashed luminosity function arises mainly from the redshift
dependence of $f_{HII}$ prior to the reionization. 
Thus accurately measuring the luminosity function of galaxies at these
epochs will give an independent constraint on the epoch and nature 
of reionization.

\section{Redshift evolution of the SFR density}
\begin{figure}
\centerline{
\psfig{figure=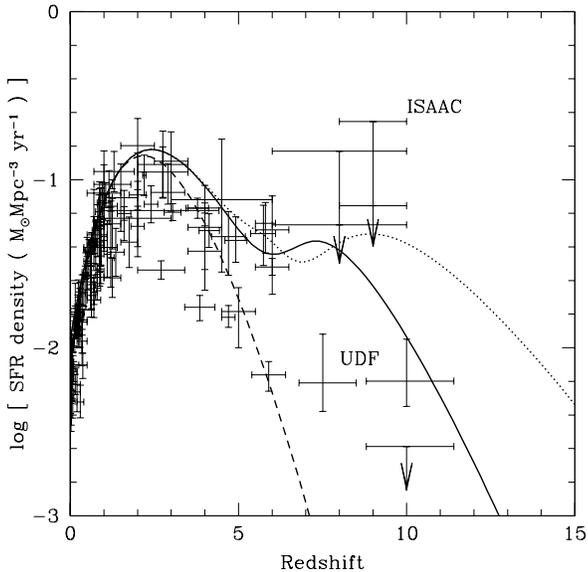,width=8cm,angle=0.0}}
\caption[]{Redshift evolution of SFR density. The observed data points
are a compendium of all the available observations obtained by Hopkins \& Beacom
(2006) and scaled down by a factor of $2.5$ to make it consistent with the IMF
used in our models.
The solid and dotted lines are the total SFR density
calculated from Eq.~(\ref{eqnsfr}) for model A and M$_2$ respectively.
The dashed line is the SFR density obtained
by integrating the luminosity function up to $0.3L^*_{z=3}$ for model A.
}
\label{figsfr0to6}
\end{figure}

Most of the semi-analytic models in the literature  use the observed star formation 
rate density to fix model parameters. In this section we discuss the evolution of
SFR density in detail in the frame work of models discussed here. The SFR density
and its redshift evolution in our model is given by Eq.~(\ref{eqnsfr}). However,
observationally,
one determines only the luminosity function above some luminosity threshold.
The SFR density is then estimated by integrating the luminosity function and
using continuous star formation with a set of IMFs. In Fig.~\ref{figsfr0to6}
we show the observed SFR density obtained by Hopkins \& Beacom (2006)
from the compendium of all the available observations and 
our predictions for a few models.
We have also added the results of Richard et al. (2006) and Bouwens et al. (2005) at
$z\ge6$ with a reddening correction of $\eta=4.5$.

First we consider our fiducial model A. 
Recall that this model adopts a Salpeter IMF with $1-100~M_\odot$.
This leads to a UV luminosity a factor $2.5$ larger than 
a Salpeter IMF with $0.1-100~M_\odot$, canonically used
to calculate the SFR density from the observed luminosity function.
We have therefore scaled down the observed data points given by
Hopkins \& Beacom (2006) appropriately.
The continuous curve in this figure is SFR density given by  Eq.~(\ref{eqnsfr}) for 
our model A.  This curve fits the observed SFR density for $z\le6$ and has a second
peak at $z\sim7.5$ before declining with increasing redshift. Such  a high
redshift peak is predicted by semi-analytical models with photoionization feedback
(see for example: Barkana \& Loeb, 2001 and Choudhury \& Srianand 2002).
It has also been pointed out that the peak becomes more pronounced and moves
towards high-$z$ if one includes molecular cooled low mass halos in the
calculations. This can be seen from the
dotted curve in the figure which shows the result for model M$_2$.
Thus if one considers only the global SFR density, it appears that the existence of 
the second peak at high-$z$  could explain the Richard et al's
point in this figure.
However, as we have already mentioned the observed SFR density is obtained by 
integrating the luminosity function up to $0.3L^*_{z=3}$. In Fig.~\ref{figsfr0to6} 
we have also shown the SFR density calculated with this prescription (dashed line)
for model A. The corresponding behavior for model M$_2$ is similar
to this dashed curve. 
It is clear that with this prescription the SFR density is always less than
that computed from Eq.~(\ref{eqnsfr}).
The difference become more and more as one moves towards the higher redshifts
as the number of low mass halos increases with the increasing redshift.
Clearly the second peak that is visible in the solid curve disappears when we
use the low luminosity cutoff while computing the SFR density.  Consistent
with our discussions on the luminosity function, our models that fit 
$z<6$ luminosity functions will not be able to explain SFR density obtained
from the Richard et al's observations.

Therefore, even though semi-analytical models predict
the SFR density in a simple analytic form,
in order to compare with the observations it is important to 
model the 
luminosity function as we have done here. Since observations
are not very sensitive to the star formation
activities in the low mass halos the measurement of SFR density directly from the 
observations will grossly under predict the actual star formation rate density 
especially at high-$z$. In this regard redshift distribution GRBs will provide a
very useful probe 
of the star formation rate density at $z\ge6$, if they trace the underlying 
star formation rates (Barkana \& Loeb, 2001; Choudhury \& Srianand 2002).

\section{Discussions and Conclusions}

We have presented here a semi-analytic formalism for computing 
(i) star formation, (ii) reionization
(iii) UV luminosity functions and (iv) source counts using a 
modified PS formalism, 
taking into account the cooling constraints, radiative feedback and suppression of
star formation in high mass halos. 

We find that even if star formation is hosted only in large atomic cooled halos
the universe is sufficiently
reionized to be consistent with the $\tau_e = 0.09\pm 0.03$ inferred 
by WMAP 3rd year data, for a range of star formation scenarios.
Also the inferred reionization redshifts are consistent with observations of
the highest redshift QSOs (Fan et al. 2006) and Lyman-$\alpha$ emitters (Iye et al. 2006).
The inclusion of star formation in molecular cooled
halos increases the value of $\tau_e$. However, the recent WMAP 
observations are better consistent with a low efficiency of the molecular cooled
halos in reionizing the universe.

The major focus of our work here is on a self-consistent modelling of 
the observed UV luminosity function of galaxies, its evolution at high redshifts and
then using this to probe the nature and evolution of star
formation at high-$z$. Our semi-analytic models with the best fit cosmological parameters derived 
from WMAP 3rd year data
fit the observed galaxy luminosity functions in the redshift range $3 \le z \le 6 $,
for a reasonable range of model parameters. 
The feedback due to photoionization is sufficient to explain the 
$z=6$ luminosity function. However we need additional feedback, 
possibly due to supernova that suppresses star formation activities in halos with
$10^{10} \lesssim (M/M_\odot) \lesssim 10^{11}$,  to explain the low luminosity end at 
$z=3$. 
Also the observed evolution of luminosity functions from $z=6$ to $z=3$ can easily be
explained with a modest change in $m_{\rm low}$ of the Salpeter IMF,
or the amount of dust reddening as 
expected from galaxy evolution.  We require roughly 50\% of the baryonic mass
to go through star formation over few dynamical time-scales. This is consistent 
with the median gas fraction of 50\% and the corresponding stellar mass inferred 
from the high-$z$ Lyman break galaxies \cite{erb06}. 
The lensing measurements of Mandelbaum et al (2006) constrain 
the mean conversion efficiency of baryons to stars to be about $20\%$,
with considerable error in some of their determinations (see their Table 3
and Figure 4). Note that upto about $35\%$ of the mass going into stars, 
is returned back to the ISM by supernovae, for the IMF as in our Model A. 
Therefore $f_*=0.5$ which we have adopted 
is not greatly in excess of even the above mean value. 
It is possible to have a smaller fraction of the baryons in a halo going into stars,
by lowering $\kappa$, or by adopting a higher $\sigma_8$ or $n_s$ than the fiducial
values favored by the WMAP 3rd year data. The first possibility is however
disfavored by the observations of Eyles et al (2007), which constrain the
age of stellar populations in high-$z$ galaxies.

The models that fit the luminosity function for $z\le6$ are consistent with
the upper limits on the luminosity functions for  $z=7$ obtained by Mannucci et al. (2007)
and the integrated source counts obtained by Bouwens et al. (2005) for $8\le z\le12$.
The observed decline in the luminosity function with increasing $z$ is naturally produced
by the decline in the halo number density coming from structure formation models
without any additional dramatic changes in the mode of star formation. 
However, if the three candidate galaxies tentatively identified by
Bouwens et al. (2005) become confirmed as high-$z$ galaxies
then we required additional changes in the mode of star formation.
Moreover, the average luminosity function obtained by Richard et al. (2006) for $6\le z\le 10$
can only be understood if star formation occurs in a 
burst mode with high efficiency,
top-heavy IMF and very little or no reddening correction for the UV light. 
These models produce more number of galaxies than the three obtained by
Bouwens et al (2005).
The difference between the two sets of available observations at $z>6$ 
is perhaps much larger than the expected cosmic variance. 
Thus a convergence from the observational front
is needed before we can draw any interesting conclusions 
on the nature of the star formation activities at $z\ge6$. 
An important constraint arises from the fact that 
the rest UV-optical spectral energy distribution of
considerable fraction of $z\sim 6$ galaxies show a Balmer break.
This suggests prolonged star formation activities with considerable 
mass contributed by low and intermediate mass stars, at least in these 
high redshift galaxies (Eyles et al. 2007).

The abundance of low luminosity galaxies is quite
sensitive to the photoionization feedback, and hence to the reionization history.
Using a range of self-consistent reionization models we show that 
such a feedback can lead to a flattening or break in the high-$z$ galaxy 
luminosity function at low luminosities 
($L(1500) \lesssim 10^{41}$ erg s$^{-1}$\AA$^{-1}$).
Accurately measured luminosity functions in the redshift range $6\le z\le 10$
can therefore be used to place interesting constraints on the epoch of reionization and
the nature of star formation activities in the dark ages. 

We compare our model predictions of star formation rate density with the
observations. The observed SFR density is obtained by integrating the 
luminosity function up to some low luminosity limit ($0.3L^*_{z=3}$).  This approach clearly
under predicts the actual global star formation rate density especially at higher redshifts.
As molecular cooled halos are expected to be fainter than the typical
detection limits achieved in the deep field images, the star formation history
constructed directly from the observations will miss any peak in the 
SFR density mainly due to such halos. In that case other tracers of
star formation activities like GRBs will be much more useful in 
detecting the enhanced star formation activities in such low mass halos.

We have used here the modified Press-Schecter formalism of Sasaki (1994),
to calculate the formation rate and survival probability of dark matter halos.
Note that simply taking the time derivative of the PS 
or some other mass function, does not give the formation rate of halo,
but only the formation minus the destruction rate.
It would be interesting to obtain the formation rate
directly from N-body simulations and repeat our calculations of
high-$z$ galaxy luminosity functions.
Metallicity of the gas is one of the
factors which could decide the nature of the stellar IMF (cf. Schneider et al 2006). Very low metallicity
could favor a top-heavy IMF, while as the metallicity increases one
may transit to a more standard Salpeter IMF. 
However, the time-scale over which the the metal enriched gas mixes with 
primordial gas is still a subject of debate; Jimenez and Haiman (2006) 
in fact point to 
observational evidence for top-heavy ``primordial'' star formation even at $z\sim 3$.
In our present models we have not explicitly included such metallicity feedback.
Although the redshift evolution of $m_{\rm low}$ required
to fit the luminosity function between $z=3$ and $z=10$ in our models,
is perhaps an indication of such a feedback. 
Our model calculations also do not consider the influence of outflows 
in suppressing star formation. Such 
outflows are important in enriching the IGM at high-$z$. 
And they could also play a role in providing the additional feedback
that we clearly need in the case of $z=3$ luminosity functions.
We hope to return to some of these issues in more detail in 
future work.

\section*{acknowledgements}
We thank Iwata Ikuru and Johan Richard for kindly providing the data on 
luminosity function used here at $z<6$ and $z>6$ respectively.
We also thank Andrew Hopkins for kindly providing the data on SFR density
and Jasjeet Bagla for discussions. We thank an anonymous referee for insightful comments.
SS thanks CSIR for providing
support for this work.

\end{document}